\documentclass[11pt,reqno]{article}
\usepackage{amsmath}
\usepackage{amsthm}
\usepackage{amssymb}
\usepackage{amsfonts}
\usepackage{mathtools}
\usepackage{cases}
\usepackage{graphicx}
\usepackage{xcolor}
\usepackage[margin=1in]{geometry}
\usepackage[utf8]{inputenc}
\usepackage[margin=1in]{geometry}
\usepackage{verbatim}
\usepackage{caption,subcaption}
\usepackage{algorithm}
\usepackage{algpseudocode}
\usepackage[normalem]{ulem}
\makeatletter
\renewcommand{\fnum@algorithm}{\fname@algorithm}
\makeatother
\usepackage[colorlinks,citecolor=blue,urlcolor=blue]{hyperref}
\usepackage{bm}
\usepackage{multirow}
\usepackage{makecell}
\usepackage{adjustbox} 
\usepackage{tabu,multirow}
\usepackage{multirow, multicol, makecell, booktabs} 
 \usepackage{mathrsfs}
 
 \usepackage{booktabs}
\usepackage{microtype}
\usepackage{enumitem}
\usepackage[T1]{fontenc}
\usepackage[utf8]{inputenc}
\numberwithin{equation}{section}
\newtheorem{definition}{Definition}[section]
\newtheorem{remark}{Remark}[section]
\newtheorem{theorem}{Theorem}[section]
\newtheorem{lemma}{Lemma}[section]

\newcommand{\be}{\begin{equation}}
\newcommand{\ee}{\end{equation}}
\newcommand{\bee}{\begin{equation*}}
\newcommand{\eee}{\end{equation*}}
\newcommand{\bi}{\begin{itemize}}
\newcommand{\ei}{\end{itemize}}
\DeclareMathOperator{\tr}{tr}

\usepackage{algorithm}
\usepackage{algpseudocode}
\usepackage{soul}%delete line

\def \N{\mathbb{N}}

\def \R{\mathbb{R}}

\newcommand{\cL}{\mathcal{L} }

\newcommand{\ea}{\end{eqnarray}}
\newcommand{\ba}{\begin{eqnarray}}
\newcommand{\ean}{\end{eqnarray*}}
\newcommand{\ban}{\begin{eqnarray*}}

\newcommand{\en}{\end{equation}}

\usepackage{graphicx}
\graphicspath{{./fig/}}

\usepackage{bm}
\usepackage{natbib}

\usepackage{amsmath,amssymb,amsthm,mathtools}
\usepackage{bm}
\usepackage{microtype}
\usepackage{enumitem}
\usepackage{setspace}
\usepackage{hyperref}
\newcommand{\vecop}{\operatorname{vec}}

\newcommand{\T}{\mathsf{T}}

\newcommand{\B}{\mathcal{B}}
\newcommand{\Law}{\mathcal{L}}
\newcommand{\Id}{\operatorname{Id}}
\newcommand{\unif}{\operatorname{Unif}}
\title{Vector Copula-Based Markov Models}
\author{
Li-Hsien Sun\thanks{
           Graduate Institute of Statistics, National Central University, Taiwan, email:\texttt{lihsiensun@ncu.edu.tw}. Partially supported by National Science and Technology Council (NSTC 115-2118-M-008-003-MY2 and NSTC 115-2123-M-035-001)
}
\and 
Po-Chun Huang\thanks{
Department of Mathematics, National Tsing Hua University, Taiwan
}
\and 
Po-Cheng Chen\thanks{
           Graduate Institute of Statistics, National Central University, Taiwan
           }
 \and
Chi-Yang Chiu\thanks{ Division of Biostatistics, Department of Preventive Medicine, University of Tennessee Health Science Center, Memphis, Tennessee, USA}     
}

\begin{document}
\maketitle

\begin{abstract}
We propose the vector copula-based Markov model (VCBMM) for describing multivariate time series with nonlinear serial dependence. To extend univariate copula-based Markov models to multivariate time series, we develop a hierarchical construction based on scalar aggregate variables and consider the vector Sklar theorem to obtain the joint density of the VCBMM. We further assume the Clayton copula for serial dependence and the multivariate Gaussian distribution for the marginals. The corresponding likelihood function and maximum likelihood estimation (MLE) with the inference functions for margins (IFM) method are provided. The performance of the proposed model is illustrated through simulation studies. In real data analysis, we apply the model to bivariate daily log returns of stocks from the 2008 global financial crisis (2008 GFC) and the 2020 COVID-19 pandemic (2020 COVID-19).

%This thesis introduces a novel modeling framework, the Hierarchical Vector Copula-based first-order Markov process, designed to capture complex dependencies in multidimensional time series data. Traditional copula models often face computational bottlenecks or restrictive assumptions when applied to standard multivariate distributions. To overcome these limitations, we propose a generalized methodology that effectively decouples cross-sectional (spatial) correlations from temporal dependencies. By utilizing mathematical transformation techniques to map dependent variables into independent components, we derive an exact and tractable formulation for the joint density function without relying on restrictive classical forms. For model estimation, we employ a robust hybrid strategy combining Maximum Likelihood Estimation with Inference Functions for Margins to ensure both computational stability and asymptotic efficiency. The mathematical mechanics and practical applicability of the proposed framework are validated through comprehensive simulation studies and empirical analysis of global financial crises. Ultimately, our findings demonstrate that this new architecture provides a robust, flexible, and promising solution for characterizing asymmetric risk contagion and extreme tail movements in multidimensional data.

% %
%%
%%
%%  
%% 
%%
%%
%% 
%%
%%
%%
\end{abstract}


\textbf{MSC (2020):} 
62M10, %Time series, auto-correlation, regression, etc. in statistics (GARCH)
60G07, %General theory of stochastic processes
62F40  %Point estimation
%91G45, %Financial networks (including contagion, systemic risk, regulation)
%91A80, %Applications of game theory
%93E11 %Filtering in stochastic control theory [See also 60G35]
%91G80.
%60G40,  	%Stopping times; optimal stopping problems; gambling theory
%60J45,  	%Probabilistic potential theory
%93E20,  % Optimal stochastic control
%60H10, %Stochastic ordinary differential equations
%94A17 %Measures of information, entropy
%91A13, % Games with infinitely many players
%49K21. % Optimality conditions:	Problems involving relations other than differential equations
%60J05,  %	Discrete-time Markov processes on general state spaces
\smallskip

\section{Introduction}
Multivariate time series models describe dependence within each observation vector and across time. Vector autoregressive (VAR) models provide a standard approach to this task (\cite{sims1980macroeconomics, lutkepohl2005new, tsay2013multivariate,DMTW2025}). A Gaussian VAR specifies linear conditional-mean dynamics. For applications involving nonlinear or asymmetric serial dependence, copula-based transitions provide an alternative way to specify dependence while retaining a chosen marginal distribution. We consider this approach for observation vectors with a prescribed multivariate marginal distribution.

% To analyze the multivariate time series analysis, in the literature, vector autoregressive (VAR) models are
%standard tools for analyzing multivariate time series  . While the VAR model successfully extends the analysis to multiple dimensions,  the standard Gaussian VAR models linear conditional-mean dynamics and does not directly parameterize asymmetric tail dependence. This motivates alternative dependence models.
%See \cite{choros2010, joe2014} for instance. Consequently, it is interesting to develop a rigorous statistical framework capable of modeling multivariate vector time series while explicitly accommodating complex, nonlinear dependence structures. 

To capture nonlinear dependence, copulas separate marginal distributions from their dependence
structure \cite{sklar1959}. For sequential data analysis, the copula-based Markov model (CBMM) introduced by \cite{CopulaMarkov} can be considered. According to \cite{CWY2009}, there are several advantages of the CBMM. First, a time-homogeneous copula-based transition can preserve a
specified invariant marginal distribution. Initializing the chain from that distribution yields a
strictly stationary process. Second, the flexibility of choices of copulas and marginal distributions allows the copula and marginal distribution to be selected to suit
the data. Theoretically, \cite{CF2006b} discuss the model selection of semiparametric copula based models under copula misspecification. \cite{CWY2009} propose the efficient estimation of copula-based Markov chain models. \cite{CF2006a,DGP2009,Markov02, Markov03,Huang2021,Joe1997,Kim2021,Li2019,SHAKE2020,Zhang2021} consider copula-based Markov chain models for a time series.

Copula constructions provide ways to model dependence between multivariate observation vectors while preserving specified marginal distributions. \cite{brechmann2014} develops a general hierarchical Kendall construction and gives particular attention to inference with Archimedean component copulas, for
which analytical expressions are available. \cite{FH2023} achieve a major breakthrough by introducing a generalized vector copula theory utilizing measure transport tools to handle complex vector dependence. 
\cite{SD2026} develop a vector-vine framework for multivariate longitudinal data that accommodates finite-order Markov dependence and stationary processes. Building on these constructions, we study a stationary first-order vector copula-based Markov model (VCBMM) in which serial dependence is mediated by a scalar aggregate of each observation vector. The aggregate is uniform under the specified marginal model, and a bivariate copula links consecutive aggregates. This specification assumes that previous vectors with the same aggregate produce the same predictive distribution for the next vector. For the Gaussian–Clayton specification, we construct the aggregate using the Cholesky whitening and the Kendall distribution transformation. A Clayton copula represents lower-tail serial dependence between consecutive aggregates. The simulations and empirical applications consider bivariate Gaussian marginals. We formulate the full likelihood and derive analytical score and Hessian expressions for the Gaussian–Clayton specification. Pa

Parameters are estimated by maximum likelihood estimation (MLE) and the inference functions for margins (IFM) estimates  introduced in \cite{FH2023} for comparison. The full likelihood accounts for the dependence of the aggregate on the marginal parameters. Simulations examine finite-sample estimation performance across sample sizes, cross-sectional correlations, and serial dependence strengths. The MLE and the IFM perform similarly under weak serial dependence, while maximum likelihood generally has smaller root mean squared errors under stronger dependence in the investigated settings. Applications to bivariate daily log returns of stocks from the 2008 global financial crisis and the 2020 COVID-19 pandemic illustrate estimation of serial dependence between aggregate variables.

% Hence, to analyze the multivariate time series with nonlinear serial dependence, we construct the vector copula-based Markov model (VCBMM). We further assume that the serial dependence follows the Clayton copula focusing on the lower tail dependence which is important for investment or hedging in financial applications. For the simulations and empirical applications, we specify a
%bivariate Gaussian marginal distribution. Furthermore, we introduce the likelihood function for the proposed VCBMM. Regarding the estimation, we develop the maximum likelihood estimation (MLE) with the inference functions for margins (IFM) method. In simulation studies, we validate its finite-sample properties through the Monte Carlo method especially 
%focusing on its performance under extreme parameters and limited sample sizes. In the empirical analysis, we illustrate the model using bivariate daily log returns during the 2008 global financial crisis (2008 GFC) and the 2020 pandemic-period (2020 COVID-19) sample.

The remainder of this paper is organized as follows. Section \ref{MR} presents the model construction, likelihood, and estimation procedure. Section \ref{SS} reports the simulation results, and Section \ref{ES} presents the financial applications. Section \ref{CON} concludes and outlines possible extensions.

% The remainder of this paper is organized as follows. In Section \ref{MR}, we propose the vector copula-based Markov model and the corresponding likelihood function and estimation. The performance of the proposed method is illustrated in Section \ref{SS}. We also provide the real data analysis using the multivariate daily returns during the 2008 GFC and 2020 COVID-19 in Section \ref{ES}. Section \ref{CON} concludes the paper. 

 \section{Main Results}\label{MR}

In this section, we rigorously derive the joint probability density function for the proposed VCBMM. The derivation combines the theoretical framework of vector copulas with a hierarchical construction designed to capture complex dependence structures.

\subsection{Vector Copula-Based Markov Models}

We begin with a formal characterization of the joint distribution. Following the framework introduced by \cite{FH2023}, the notion of a vector copula is defined as follows.

\begin{definition}[Vector Copula]
Let $K\in\N$ and let $d_1,\ldots,d_K\in\N$ be positive integers. Set
\[
 d=\sum_{k=1}^K d_k.
\]
A \emph{vector copula} $C$ is a probability distribution on $[0,1]^d$ such that, for each $k=1,\ldots,K$, the corresponding nonoverlapping $d_k$-dimensional marginal distribution is uniform on $[0,1]^{d_k}$.
\end{definition}

Accordingly, the following result gives a transport-based characterization of a vector copula.

\begin{theorem}[Vector Sklar Theorem]\label{thm:vector-sklar}
Let $d_1,\ldots,d_K$ be positive integers and define
\[
 \Pi_k=[0,1]^{d_k},\qquad k=1,\ldots,K.
\]
Let $P$ be a probability measure on $\R^{d_1}\times\cdots\times\R^{d_K}$ with marginal distributions $P_k$ on $\R^{d_k}$. For each $k$, let $\psi_k$ be a convex potential such that the gradient map $\nabla\psi_k$ transports the uniform probability measure $\mu_k$ on $\Pi_k$ to $P_k$, that is,
\[
 (\nabla\psi_k)_{\#}\mu_k=P_k.
\]
Assume that each $P_k$ is absolutely continuous with respect to Lebesgue measure on $\R^{d_k}$ and that its support is contained in a convex set. Then there exists a unique vector copula $C$, with associated probability measure $P_C$ on $\Pi_1\times\cdots\times\Pi_K$, such that the following properties hold.

\begin{enumerate}[label=(\arabic*)]
\item There exists a joint probability distribution on
\[
 (\R^{d_1}\times\cdots\times\R^{d_K})\times(\Pi_1\times\cdots\times\Pi_K)
\]
whose marginals are $P$ and $P_C$, respectively, and whose $(X_k,U_k)$-marginal is
\[
 (\nabla\psi_k,\Id)_{\#}\mu_k,\qquad k=1,\ldots,K.
\]

\item For any Borel sets $A_k\subseteq\R^{d_k}$,
\begin{equation}
P(A_1\times\cdots\times A_K)
=
P_C\!\left((\nabla\psi_1)^{-1}(A_1)\times\cdots\times(\nabla\psi_K)^{-1}(A_K)\right),
\label{eq:2.1}
\end{equation}
where
\[
(\nabla\psi_k)^{-1}(A_k)
:=\{u\in\Pi_k:\nabla\psi_k(u)\in A_k\}
\]
denotes the set-theoretic preimage.

\item If $\psi_k^*$ denotes the convex conjugate of $\psi_k$, then the corresponding Brenier maps satisfy
\[
 \nabla\psi_k^*\circ\nabla\psi_k=\Id
 \quad \mu_k\text{-a.e.},
 \qquad
 \nabla\psi_k\circ\nabla\psi_k^*=\Id
 \quad P_k\text{-a.e.}
\]
for $k=1,\ldots,K$. Consequently, for Borel sets $B_k\subseteq\Pi_k$,
\begin{equation}
P_C(B_1\times\cdots\times B_K)
=
P\!\left((\nabla\psi_1^*)^{-1}(B_1)\times\cdots\times(\nabla\psi_K^*)^{-1}(B_K)\right),
\label{eq:2.2}
\end{equation}
where
\[
(\nabla\psi_k^*)^{-1}(B_k)
:=\{x\in\R^{d_k}:\nabla\psi_k^*(x)\in B_k\}.
\]
\end{enumerate}
\end{theorem}

Theorem~\ref{thm:vector-sklar} shows that a joint distribution with suitable marginals induces a unique vector copula. To complete the theoretical picture, we now consider the converse direction: given a vector copula and prescribed marginal distributions, we construct a joint distribution having exactly those marginals.

\begin{theorem}[Joint Distribution Induced by a Vector Copula with Prescribed Marginals]\label{thm:forward}
Let $K\in\N$ and $d_1,\ldots,d_K\in\N$. For each $k=1,\ldots,K$, let
\[
\mu_k=\unif([0,1]^{d_k})
\]
and let $P_k$ be a Borel probability measure on $\R^{d_k}$. Let $P_C$ be a probability measure on
\[
[0,1]^{d_1}\times\cdots\times[0,1]^{d_K}
\]
whose $k$th marginal is $\mu_k$; equivalently, $P_C$ represents a $K$-block vector copula.

Suppose that, for each $k$, there exists a Borel-measurable map
\[
\mathcal{T}_k:[0,1]^{d_k}\longrightarrow\R^{d_k}
\]
such that
\[
(\mathcal{T}_k)_{\#}\mu_k=P_k.
\]
In particular, whenever $P_k$ admits a vector-quantile representation, one may take
\[
\mathcal{T}_k=\nabla\psi_k\qquad \mu_k\text{-a.e.},
\]
where $\psi_k:[0,1]^{d_k}\to\R$ is convex.

Define the product map
\[
\mathcal{T}(u_1,\ldots,u_K)
:=\bigl(\mathcal{T}_1(u_1),\ldots,\mathcal{T}_K(u_K)\bigr)
\]
and the pushforward measure
\[
P:=\mathcal{T}_{\#}P_C.
\]
Then:
\begin{enumerate}[label=(\arabic*)]
\item $P$ is a probability measure on $\R^{d_1}\times\cdots\times\R^{d_K}$ whose $k$th marginal is $P_k$.
\item For every collection of Borel sets $A_k\in\B(\R^{d_k})$,
\begin{equation}
P(A_1\times\cdots\times A_K)
=
P_C\!\left(\mathcal{T}_1^{-1}(A_1)\times\cdots\times \mathcal{T}_K^{-1}(A_K)\right),
\label{eq:2.3}
\end{equation}
where
\[
\mathcal{T}_k^{-1}(A_k):=\{u\in[0,1]^{d_k}:\mathcal{T}_k(u)\in A_k\}
\]
denotes the set-theoretic preimage. In particular, if $\mathcal{T}_k=\nabla\psi_k$, then
\begin{equation}
P(A_1\times\cdots\times A_K)
=
P_C\!\left((\nabla\psi_1)^{-1}(A_1)\times\cdots\times(\nabla\psi_K)^{-1}(A_K)\right).
\label{eq:2.4}
\end{equation}
\end{enumerate}
\end{theorem}

\begin{remark}\label{rem:preimage}
In Theorem~\ref{thm:forward}, the notation $(\nabla\psi_k)^{-1}(A_k)$ denotes the set-theoretic preimage of $A_k$ under $\nabla\psi_k$. It does not require $\nabla\psi_k$ to possess a classical pointwise inverse.
\end{remark}

\begin{proof}[Proof of Theorem~\ref{thm:forward}]
Since each $\mathcal{T}_k$ is Borel measurable, the product map
\[
\mathcal{T}=\mathcal{T}_1\times\cdots\times \mathcal{T}_K
\]
is also Borel measurable. Hence $P=\mathcal{T}_{\#}P_C$ is a well-defined probability measure.

For a measurable rectangle $A_1\times\cdots\times A_K$, the definition of a pushforward measure gives
\begin{align*}
P(A_1\times\cdots\times A_K)
&=\mathcal{T}_{\#}P_C(A_1\times\cdots\times A_K)\\
&=P_C\!\left(\mathcal{T}^{-1}(A_1\times\cdots\times A_K)\right)\\
&=P_C\!\left(\mathcal{T}_1^{-1}(A_1)\times\cdots\times \mathcal{T}_K^{-1}(A_K)\right),
\end{align*}
which proves \eqref{eq:2.3}. Equation \eqref{eq:2.4} follows immediately when $\mathcal{T}_k=\nabla\psi_k$.

It remains to verify the marginals. Fix $k\in\{1,\ldots,K\}$ and let $A_k\in\B(\R^{d_k})$. Denote by $P_k^{(P)}$ the $k$th marginal of $P$. Then
\begin{align*}
P_k^{(P)}(A_k)
&=P\bigl(\R^{d_1}\times\cdots\times A_k\times\cdots\times\R^{d_K}\bigr)\\
&=P_C\bigl([0,1]^{d_1}\times\cdots\times \mathcal{T}_k^{-1}(A_k)\times\cdots\times[0,1]^{d_K}\bigr)\\
&=\mu_k\bigl(\mathcal{T}_k^{-1}(A_k)\bigr)\\
&=(\mathcal{T}_k)_{\#}\mu_k(A_k)\\
&=P_k(A_k).
\end{align*}
Thus the $k$th marginal of $P$ is exactly $P_k$. Since $k$ was arbitrary, all prescribed marginals are recovered.
\end{proof}

Having established Theorems~\ref{thm:vector-sklar} and \ref{thm:forward}, we now clarify their relationship. Together, these results describe the forward construction and the converse recovery in the transport-based formulation of the vector version of Sklar's theorem.

\begin{remark}\label{rem:complementary}
Theorems~\ref{thm:vector-sklar} and \ref{thm:forward} may be viewed as two complementary directions of the same transport-based formulation of a vector version of Sklar's theorem.

In Theorem~\ref{thm:forward}, one starts from a $K$-block vector copula $P_C$ on
\[
[0,1]^{d_1}\times\cdots\times[0,1]^{d_K},
\]
together with measurable transport maps
\[
\mathcal{T}_k:[0,1]^{d_k}\to\R^{d_k},
\qquad
(T_k)_{\#}\mu_k=P_k.
\]
Defining
\[
\mathcal{T}(u_1,\ldots,u_K)=\bigl(\mathcal{T}_1(u_1),\ldots,\mathcal{T}_K(u_K)\bigr),
\qquad
P:=T_{\#}P_C,
\]
produces a joint distribution with the prescribed marginals $P_1,\ldots,P_K$. In particular, for Borel sets $A_k\in\B(\R^{d_k})$,
\[
P(A_1\times\cdots\times A_K)
=
P_C\!\left(\mathcal{T}_1^{-1}(A_1)\times\cdots\times \mathcal{T}_K^{-1}(A_K)\right).
\]

Conversely, under the additional assumptions of Theorem~\ref{thm:vector-sklar} that guarantee almost-everywhere inverse relations between the vector-quantile maps and the corresponding vector-rank maps, the associated vector copula can be recovered from the joint distribution $P$. More precisely, if $\mathcal{T}_k=\nabla\psi_k$ and $\psi_k^*$ denotes the convex conjugate of $\psi_k$, with
\[
\nabla\psi_k^*\circ\nabla\psi_k=\Id
\quad \mu_k\text{-a.e.},
\qquad
\nabla\psi_k\circ\nabla\psi_k^*=\Id
\quad P_k\text{-a.e.},
\]
then
\[
P_C
=
(\nabla\psi_1^*,\ldots,\nabla\psi_K^*)_{\#}P.
\]
Thus, under these almost-everywhere invertibility conditions, Theorem~\ref{thm:forward} describes the forward construction $P_C\mapsto P$, whereas Theorem~\ref{thm:vector-sklar} describes the corresponding recovery $P\mapsto P_C$.
\end{remark}

The preceding results establish the validity of a hierarchical construction for modeling dependence among multivariate random vectors through a lower-dimensional nesting copula. Specifically, each multidimensional random vector is associated with a scalar aggregate variable through a prescribed aggregation map, while cross-vector dependence is mediated by these aggregate variables. The dependence among the aggregates is then characterized by a lower-dimensional copula at the upper level of the hierarchy, whereas the intrinsic dependence structure within each multidimensional random vector is retained at the lower level. Under the regularity and hierarchical dependence assumptions stated below, the resulting construction yields a well-defined and internally consistent joint dependence structure for the original multidimensional random vectors.

\paragraph*{Hierarchical dependence structure.}
The dependence between $Y_1$ and $Y_2$ is assumed to be entirely mediated by their corresponding scalar aggregate variables
\[
V_1=T_1(Y_1),\qquad V_2=T_2(Y_2),
\]
through an absolutely continuous nesting copula $C_0$ with density $c_0$. The maps $T_k$ are chosen so that $V_k$ is uniformly distributed on $[0,1]$; for example, this may be achieved through an appropriate Kendall-distribution transformation in a hierarchical copula construction.

\begin{theorem}[Hierarchical Joint Density with Vector Copula]\label{thm:hierarchical-density}
Let $(\Omega,\mathcal F,\mathbb P)$ be a probability space, and let
\[
Y_1:\Omega\to\R^{d_1},\qquad
Y_2:\Omega\to\R^{d_2}
\]
be random vectors with absolutely continuous marginal distributions having densities $f_1$ and $f_2$, respectively. Let
\[
V_k=T_k(Y_k),\qquad k=1,2,
\]
where $T_k:\R^{d_k}\to[0,1]$ is measurable and $V_1,V_2\sim\unif([0,1])$.

Assume the following conditions:
\begin{enumerate}[label=(A\arabic*)]
\item \textbf{Conditional independence.} The random vectors $Y_1$ and $Y_2$ are conditionally independent given $V=(V_1,V_2)^{\mathsf T}$.
\item \textbf{Sufficiency of the aggregates.} For $k\in\{1,2\}$, the conditional distribution of $Y_k$ given $V$ depends only on its corresponding aggregate $V_k$, that is,
\[
\Law(Y_k\mid V)=\Law(Y_k\mid V_k).
\]
\end{enumerate}
If the dependence between $V_1$ and $V_2$ is governed by an absolutely continuous copula $C_0$ with density $c_0$, then the joint density of $(Y_1,Y_2)$ is
\begin{equation}
f(y_1,y_2)
=
c_0\!\left(T_1(y_1),T_2(y_2)\right)f_1(y_1)f_2(y_2)
\label{eq:2.5}
\end{equation}
for Lebesgue-a.e. $(y_1,y_2)\in\R^{d_1}\times\R^{d_2}$.
\end{theorem}

\begin{proof}
Let $P_k$ denote the marginal distribution of $Y_k$, $k=1,2$. Since $P_k$ is absolutely continuous with respect to Lebesgue measure on $\R^{d_k}$, it admits density $f_k$.

Because $V_1,V_2\sim\unif([0,1])$ and their dependence is described by the absolutely continuous copula $C_0$, Sklar's theorem gives
\begin{equation}
F_V(v_1,v_2)
=C_0(F_{V_1}(v_1),F_{V_2}(v_2))
=C_0(v_1,v_2),
\qquad 0\le v_1,v_2\le 1.
\label{eq:2.6}
\end{equation}
Since $C_0$ is absolutely continuous,
\[
C_0(v_1,v_2)
=\int_0^{v_1}\int_0^{v_2} c_0(s,t)\,dt\,ds,
\]
so the joint density of $V=(V_1,V_2)$ is
\[
f_V(v_1,v_2)=c_0(v_1,v_2)
\]
for Lebesgue-a.e. $(v_1,v_2)\in[0,1]^2$.

For $j=1,2$, let
\[
K_j(v_j,A_j)
:=\mathbb P(Y_j\in A_j\mid V_j=v_j),
\qquad A_j\in\B(\R^{d_j}),
\]
denote a regular conditional distribution of $Y_j$ given $V_j$. By assumption (A2),
\[
\mathbb P(Y_j\in A_j\mid V_1,V_2)
=K_j(V_j,A_j),
\qquad j=1,2.
\]
By assumption (A1),
\begin{align}
&\mathbb P(Y_1\in A_1,Y_2\in A_2\mid V_1,V_2)\notag\\
&\qquad=
\mathbb P(Y_1\in A_1\mid V_1,V_2)
\mathbb P(Y_2\in A_2\mid V_1,V_2)\notag\\
&\qquad=K_1(V_1,A_1)K_2(V_2,A_2).
\label{eq:2.7}
\end{align}

By the tower property,
\begin{align}
\mathbb P(Y_1\in A_1,Y_2\in A_2)
&=\mathbb E\!\left[
\mathbb P(Y_1\in A_1,Y_2\in A_2\mid V_1,V_2)
\right] \notag\\
&=\mathbb E\!\left[K_1(V_1,A_1)K_2(V_2,A_2)\right].
\label{eq:2.8}
\end{align}
Using the joint density $f_V(v_1,v_2)=c_0(v_1,v_2)$, we obtain
\begin{equation}
\mathbb P(Y_1\in A_1,Y_2\in A_2)
=
\int_0^1\int_0^1
K_1(v_1,A_1)K_2(v_2,A_2)c_0(v_1,v_2)\,dv_1\,dv_2.
\label{eq:2.9}
\end{equation}

The following lemma is used to transform the conditional-kernel representation into an integral with respect to the marginal densities.

\begin{lemma}\label{lem:kernel-transform}
For any nonnegative measurable function $g:[0,1]\to[0,\infty]$, any Borel set $A\in\B(\R^{d_j})$, and $j\in\{1,2\}$,
\begin{equation}
\int_0^1 K_j(v,A)g(v)\,dv
=
\int_A g(T_j(y))f_j(y)\,dy.
\label{eq:2.10}
\end{equation}
\end{lemma}

\begin{proof}[Proof of Lemma~\ref{lem:kernel-transform}]
Since $V_j\sim\unif([0,1])$, its density is $f_{V_j}(v)=1$ for $0\le v\le1$. Hence
\begin{align*}
\int_0^1 K_j(v,A)g(v)\,dv
&=\int_0^1 \mathbb P(Y_j\in A\mid V_j=v)g(v)f_{V_j}(v)\,dv\\
&=\mathbb E\!\left[g(V_j)\mathbb P(Y_j\in A\mid V_j)\right]\\
&=\mathbb E\!\left[g(V_j)\mathbf 1_{\{Y_j\in A\}}\right]\\
&=\mathbb E\!\left[g(T_j(Y_j))\mathbf 1_{\{Y_j\in A\}}\right]\\
&=\int_A g(T_j(y))f_j(y)\,dy.
\end{align*}
This proves \eqref{eq:2.10}.
\end{proof}

Applying Tonelli's theorem to \eqref{eq:2.9} and using Lemma~\ref{lem:kernel-transform} twice yields
\begin{align*}
\mathbb P(Y_1\in A_1,Y_2\in A_2)
&=\int_0^1 K_1(v_1,A_1)
\left[\int_0^1 K_2(v_2,A_2)c_0(v_1,v_2)\,dv_2\right]dv_1\\
&=\int_0^1 K_1(v_1,A_1)
\left[\int_{A_2}c_0(v_1,T_2(y_2))f_2(y_2)\,dy_2\right]dv_1\\
&=\int_{A_2}\left[
\int_0^1 K_1(v_1,A_1)c_0(v_1,T_2(y_2))\,dv_1
\right]f_2(y_2)\,dy_2\\
&=\int_{A_2}\int_{A_1}
 c_0(T_1(y_1),T_2(y_2))f_1(y_1)f_2(y_2)
\,dy_1\,dy_2.
\end{align*}
Therefore,
\begin{equation}
\mathbb P(Y_1\in A_1,Y_2\in A_2)
=
\int_{A_2}\int_{A_1}
 c_0(T_1(y_1),T_2(y_2))f_1(y_1)f_2(y_2)
\,dy_1\,dy_2.
\label{eq:2.11}
\end{equation}
Since \eqref{eq:2.11} holds for all Borel sets $A_1$ and $A_2$, the joint density is given by \eqref{eq:2.5} for Lebesgue-a.e. $(y_1,y_2)$.
\end{proof}

We now specialize the preceding construction to a strictly stationary first-order Markov process. Let $\{Y_t\}_{t=1}^K$ be a strictly stationary first-order Markov process taking values in $\R^d$, and let $f_{\mathrm{marg}}$ denote the common marginal density of $Y_t$. Define a measurable scalar aggregate transformation
\[
T:\R^d\to[0,1]
\]
according to the hierarchical construction above, and assume that $T(Y_t)\sim\unif([0,1])$. By Theorem~\ref{thm:hierarchical-density}, the serial dependence between two consecutive observations $Y_{t-1}$ and $Y_t$ may be represented by a bivariate nesting copula $C_0$ with density $c_0$. Consequently, the joint density of $(Y_{t-1},Y_t)$ admits the factorization
\begin{equation}
f(y_{t-1},y_t)
=
c_0\!\left(T(y_{t-1}),T(y_t)\right)
 f_{\mathrm{marg}}(y_{t-1})f_{\mathrm{marg}}(y_t).
\label{eq:2.12}
\end{equation}

Since the process is first-order Markov, its transition density is
\begin{equation}
f(y_t\mid y_{t-1})
=
\frac{f(y_{t-1},y_t)}{f_{\mathrm{marg}}(y_{t-1})},
\qquad
f_{\mathrm{marg}}(y_{t-1})>0.
\label{eq:2.13}
\end{equation}
Substituting \eqref{eq:2.12} into \eqref{eq:2.13} gives
\begin{align}
f(y_t\mid y_{t-1})
&=
\frac{
c_0\!\left(T(y_{t-1}),T(y_t)\right)
 f_{\mathrm{marg}}(y_{t-1})f_{\mathrm{marg}}(y_t)
}{f_{\mathrm{marg}}(y_{t-1})}\notag\\
&=
c_0\!\left(T(y_{t-1}),T(y_t)\right)
 f_{\mathrm{marg}}(y_t).
\label{eq:conditional_density}
\end{align}
Thus, the transition mechanism decomposes naturally into two components: the marginal behavior of $Y_t$, represented by $f_{\mathrm{marg}}(y_t)$, and the serial dependence between consecutive observations, represented by the copula density $c_0$. In particular, the copula term modulates the marginal density according to the dependence between the aggregate variables $T(Y_{t-1})$ and $T(Y_t)$.

\subsection{Likelihood Function and Estimation}
For the observed sequential data $( {y}_1, \dots,  {y}_K)$, the log-likelihood function $\mathcal{L}(\Theta)$ can be decomposed into the sum of the unconditional log-likelihood of the first observation and the conditional log-likelihoods of the subsequent observations. Substituting the derived transition density from \eqref{eq:conditional_density}, we obtain
\begin{equation}
	\begin{aligned}
		\mathcal{L}(\Theta) &= \log f_{\text{marg}}( {y}_1; \theta_{\text{marg}}) + \sum_{t=2}^K \log f( {y}_t |  {y}_{t-1}; \Theta) \\
		&= \sum_{t=1}^K \log f_{\text{marg}}({y}_t; \theta_{\text{marg}}) + \sum_{t=2}^K \log c_0(V( {y}_{t-1}), V({y}_t); \theta_{c_0})~,
	\end{aligned}
	\label{eq:full_likelihood}
\end{equation}

\noindent where $\Theta = (\theta_{\text{marg}}, \theta_{c_0})$ collects all model parameters. Note that $\theta_{\text{marg}}$ includes parameters for the marginal distributions such as the mean and covariance for the Gaussian distribution defining the aggregate variable $V(\cdot)$ introduced in Section \ref{Aggregate}.  

To ensure statistical efficiency, we employ the MLE method. The estimator is defined as:
\begin{equation}
	\widehat{\Theta}_{\text{MLE}} = \operatorname*{argmax}_{\Theta} \mathcal{L}(\Theta)~.
\end{equation}

%Since the proposed bivariate model ($d=2$) allows for feasible numerical optimization, maximizing the joint likelihood function ensures that the interaction between the marginal parameters and the dependence parameter $\alpha$ is fully accounted for.

However, the hierarchical nature of the objective function can present challenges with multiple local maxima. To address this, we also consider the IFM method in \cite{FH2023} solely as a robust initialization strategy:
\begin{enumerate}
	\item \textbf{Step 1 (Initialization):} We first estimate the marginal parameters $\theta_{\text{marg}}$ independently, and then estimate $\theta_{c_0}$ given these fixed marginals. This computationally cheap two-step procedure provides a consistent (though inefficient) starting point $\Theta^{(0)}$.
	\item \textbf{Step 2 (Full Optimization):} We then use $\Theta^{(0)}$ as the initial value for the full MLE optimization in Equation \eqref{eq:full_likelihood}.
\end{enumerate}

\noindent This hybrid approach leverages the computational stability of the IFM to guide the optimizer, while ultimately delivering the asymptotic efficiency of the MLE.

In the proposed model, for simplicity, we assume that the cross-sectional behavior of the random vector $ {Y}_t \in \mathbb{R}^d$ follows a multivariate Gaussian distribution as follows:
\begin{equation}
	 {Y}_t \sim \mathcal{N}_d( {\mu},  {\Sigma})~,
\end{equation}
where $ {\mu}$ is the mean vector and $ {\Sigma}$ is the covariance matrix. The Gaussian specification is chosen for its tractability and its ability to represent cross-sectional correlations. To model serial dependence between $V_{t-1}$ and $V_t$, we employ the {Clayton copula}, a member of the Archimedean family. The copula density is given by
\begin{equation}
	c_0(u, v; \alpha) = (1+\alpha)(uv)^{-1-\alpha}(-1 + u^{-\alpha} + v^{-\alpha})^{-2-1/\alpha}~,
\end{equation}
with the dependence parameter $\alpha >0$ for capturing the positive lower tail dependence. Note the Clayton copula only emphasize lower-tail dependence coefficient $\lambda_L=2^{-1/\alpha}$ which is suitable for analyzing the lower tail risk such as the value at risk in financial time series. The strength of dependence between successive aggregates increases with $\alpha$. For $\alpha>0$, the model allows lower-tail dependence between consecutive aggregate variables. An
application-specific interpretation of low scores requires additional justification. The corresponding likelihood function and derivatives for the MLE are obtained in Appendix \ref{MLE}.

\subsection{Aggregate Variables}\label{Aggregate}

We obtain the aggregate variable $V_t$ by applying the composition measure transports (composition MT) proposed by \citet{FH2023} by the following steps: 

%A critical challenge in applying Brechmann's hierarchical framework to Gaussian marginals is that the Kendall distribution function $K(\cdot)$ for a Gaussian copula has no closed-form expression. It implies that the aggregate variable $V_t$ can not be obtained in the traditional framework. The issue can be overcome by utilizing the Composition MT of \citet{FH2023}. The procedure consists of two steps:

\noindent \textbf{Step 1: Whitening via Composition MT}

We first apply a specific composition MT that maps the correlated vector ${Y}_t$ to a vector of transformed variables, denoted as $\widetilde{{U}}_t$. In the case of the multivariate Gaussian distribution, we consider the inverse Cholesky transformation. 
%This linear operation efficiently transforms the correlated residuals into independent inputs for the copula structure.

Let ${\Sigma} = {L}{L}^T$, where $\mathrm{L}$ is the lower triangular Cholesky factor given by 
\begin{equation}
	 {Z}_t = {L}^{-1}({Y}_t -  {\mu})~.
\end{equation}
Since ${Y}_t$ is multivariate normal, ${Z}_t$ follows a standard multivariate normal distribution with the identity covariance matrix, $\mathcal{N}_d(\mathrm{0}, \mathrm{I})$. We then apply the univariate standard normal CDF, $\Phi(\cdot)$, to describe $\widetilde{U}_t$ 
\begin{equation}\label{tildeU}
	\widetilde{{U}}_t = (\Phi(Z_{t,1}), \dots, \Phi(Z_{t,d}))^T~,
\end{equation}
where the transformed vector $\widetilde{{U}}_t$ follows the standard multivariate uniform distribution on $[0,1]^d$. 
Note that the components of the transformed vector are independent.

\noindent \textbf{Step 2: Calculating the Aggregate Variable}

\citet{brechmann2014} defines the aggregate variable $V_t$ as the probability integral transform of the copula. 
%Since the components of our transformed vector $\widetilde{{U}}_t$ are independent, the underlying copula is simply the independence copula, given by $C_{\perp}({u}) = \prod_{i=1}^d u_i$.
The Kendall distribution function for the independence copula, denoted $K_{\perp}$, has a well-known analytical form given by  
\begin{equation}\label{Kendall}
	K_{\perp}(t) = t \sum_{j=0}^{d-1} \frac{(-\ln t)^j}{j!}~.
\end{equation}
Therefore, the scalar aggregate $V_t$ can be obtained using the transformed vector written as
\begin{equation}
	V_t = K_{\perp}\left( \prod_{i=1}^d \widetilde{U}_{t,i} \right)~.
\end{equation}
where $\widetilde{U}_{t,i}=\Phi(Z_{t,i})$ is given by \eqref{tildeU}. 

%Furthermore, this aggregation step elegantly mirrors the stochastic representation introduced earlier. Since the transformed components can be geometrically represented as $\widetilde{U}_{t,i} = \exp(-R_t S_i)$, and the components of the unit simplex sum to one ($\sum_{i=1}^d S_i = 1$), the copula product simplifies exactly to the exponential of the negative radial variable:
%\begin{equation}
%	\prod_{i=1}^d \widetilde{U}_{t,i} = \exp\left(-R_t \sum_{i=1}^d S_i\right) = \exp(-R_t)~.
%\end{equation}
%Consequently, the aggregate variable can be equivalently expressed in terms of the radial energy:
%\begin{equation}
%	V_t = K_{\perp}\left( \exp(-R_t) \right)~.
%\end{equation}
%This profound equivalence directly establishes the bijective relationship, $R_t = -\ln(K_{\perp}^{-1}(V_t))$, providing the exact mathematical foundation required for the reverse data generating process utilized in our simulation studies. 

The two-step procedure provides a computationally efficient and exact method to implement the hierarchical model for Gaussian marginals, seamlessly bridging estimation and simulation.

\section{Simulation Studies}\label{SS}
To evaluate the finite-sample performance of the proposed VCBMM and to verify the consistency of the estimation, we conduct a comprehensive simulation through the Monte Carlo method. The simulation is designed with two primary objectives: (1) to assess the robustness of the estimator across varying regimes (varying correlation and tail dependence), and (2) to examine how estimation accuracy changes with sample size. 

\subsection{Simulation Design and Data Generating Mechanism}
The data generating process (DGP) follows a hierarchical ``top-down'' approach. 
%As illustrated in Figure \ref{fig:dgp_flowchart}, 
The generation starts from the latent dynamic driver and cascades down to the observed correlated vector through three distinct layers.

First, in the latent temporal process layer, we generate the time-varying probability level $V_t$. Since the serial dependence is governed by the Clayton copula $C_0(\cdot; \alpha)$, the value of $V_t$ is generated conditional on the previous state $V_{t-1}$ using the conditional copula function given by 
\[
	V_t = C_{t|t-1}^{-1}(U_t^{\text{noise}} \mid V_{t-1}; \alpha)~,
\]
where $U_t^{\text{noise}} \sim \mathcal{U}[0,1]$ is an exogenous independent noise term. This step ensures that the latent process exhibits the desired autoregressive lower-tail dependence.

Second, we map the scalar $V_t$ to an intermediate vector $\widetilde{{U}}_t$, serving as an independent bridge. Referred to 
the stochastic representation algorithm for the independence copula proposed by \citep{mcneil2009}, the vector $\widetilde{{U}}_t=(\widetilde{U}_{t,1},\cdots, \widetilde{U}_{t,d})$ is generated by 
\[
	\widetilde{U}_{t,j} = \exp(-\mathcal{R}_t \cdot S_j), \quad j=1,\dots,d,
\]
where $\mathcal{R}_t = -\ln\left( K_{\perp}^{-1}(V_t) \right)~$ with $K_{\perp}$ given by \eqref{Kendall} and a random direction vector ${S} = (S_1, \cdots, S_d)^\top$ uniformly distributed on the unit simplex by generating $d$ independent standard exponential random variables and normalizing them to sum to one.

Finally, in the spatial coloring layer, we transform the independent vector $\widetilde{{U}}_t$ into the observed vector $ {Y}_t$ by introducing the desired Gaussian correlation structure. This is achieved via the inverse composition MT:
\[
	 {Y}_t =  {\mu} + {L} \cdot \Phi^{-1}(\widetilde{ {U}}_t)~,
\]
where $L$ is the Cholesky factor of the covariance matrix $ {\Sigma}$.

\subsection{Simulation Results}

To evaluate the performance of the proposed model, we configure a baseline parameter space characterized by elevated marginal volatility and strong cross-sectional correlation. Specifically, the mean vector is set to ${\mu} = (-0.55, -0.1)^\top$, and the covariance matrix is constructed using high marginal variances ($\Sigma_{11} = 55$, $\Sigma_{22} = 28$) coupled with varying correlation coefficients $\rho$. This calibration ensures the simulated data generating process reflects the severe fluctuations inherent in extreme empirical regimes, providing a rigorous test for the model's structural robustness. 

\begin{figure}[H]
	\centering
	\includegraphics[width=0.85\textwidth]{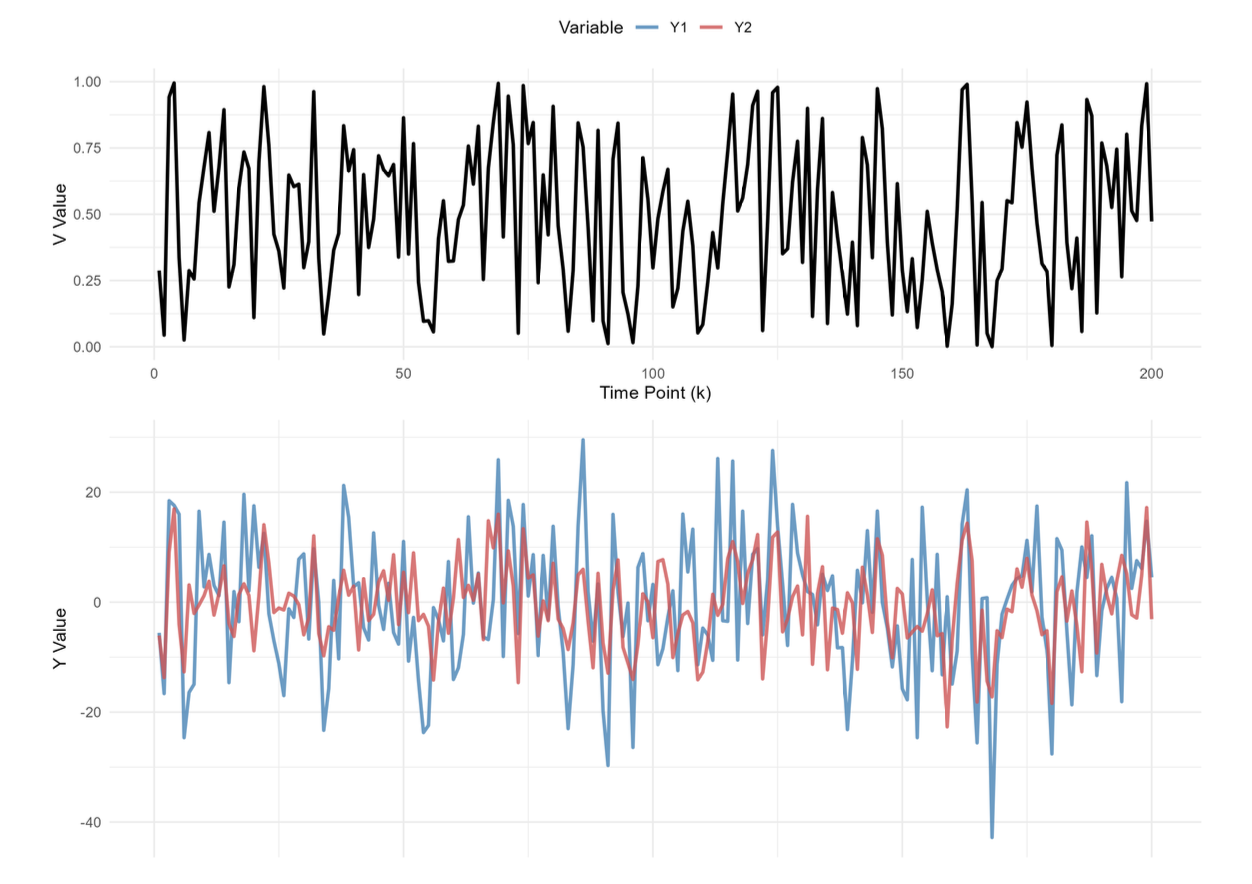} 
	\caption[Simulated Bivariate Time Series]{Comparison of aggregate variables ($V_t$) and observed ($ {Y}_t$) Time Series Dynamics. The upper panel shows the latent aggregate variable $V_t$, while the lower panel presents the observed vectors $ {Y}_t$. The simulated path demonstrates how the latent copula dynamics propagate extreme fluctuations into the observed dimensions, ensuring structural consistency between the driver and the outcome.}
	\label{fig:sim_time_series}
\end{figure}

Figure \ref{fig:sim_time_series} illustrates one realization of the aggregate and observation
processes. We next assess estimation performance by Monte Carlo simulation. The trajectory is generated using a representative parameter configuration ($\alpha=0.14, \rho=0.4$) with marginal variances calibrated to reflect a high-volatility regime. The upper panel displays the path of the latent aggregate variable $V_t$, while the lower panel tracks the resulting observed vectors ${Y}_t$. 
%The comparison illustrates the model's structural efficiency: fluctuations in the latent systematic driver $V_t$ are effectively propagated into the observed dimensions, inducing synchronous extreme movements across the components of the vector. 
This path highlights how the proposed framework captures complex joint dynamics by mapping a single temporal driver to a multidimensional observed space.

Having visually confirmed that the generative process effectively reproduces the intended dependence features, we proceed to a comprehensive numerical assessment of the MLE. The evaluation is conducted through three complementary scenarios. We use $M=2000$ Monte Carlo replications.

In the first scenario, we fix the sample size at $K=1000$ and assess the estimator across nine distinct dependence structures ($\rho \in \{-0.4, 0.1, 0.4\}$, $\alpha \in \{0.1, 2, 5\}$). Table \ref{tab:scenario_a_combined} reports mean estimates and root mean square errors (RMSEs). The findings confirm that the VCBMM framework successfully decouples the cross-sectional correlation from the latent temporal tail dependence. 
%Notably, the estimation precision of the spatial parameters remains highly stable regardless of the serial dependence intensity, validating the efficacy of the Inverse Composition MT mechanism.
We examine the estimation performance of the MLE across increasing sample sizes ($K \in \{100, 250, 500, 750, 1000\}$) under a fixed dependence structure ($\rho=0.4, \alpha=0.1$). As summarized in Table \ref{tab:scenario_b_convergence}, the RMSEs systematically diminish as the data volume expands. This systematic reduction demonstrates that the hierarchical optimization remains reliable and efficient even within highly volatile and strongly dependent parameter spaces.

Finally, the third scenario aims to bridge the simulations with real applications. By configuring the data generating process with parameters and exact finite sample lengths ($K=165$ and $K=167$) calibrated directly from two specific periods---the 2008 Global Financial Crisis and the 2020 COVID-19 pandemic---we rigorously pressure-test the estimator to ensure its practical reliability under severe, short-lived market turbulence.

%, the fourth scenario introduces a misspecification test to evaluate the estimator's structural robustness against unmodeled heavy-tailed phenomena. We simulate data using a multivariate Student-$t$ distribution with varying degrees of freedom ($\nu \in \{10, 30, 50, 100\}$), while strictly performing the estimation under a Gaussian-based likelihood. This test is designed to demonstrate the estimator's resilience to mild departures from normality, confirming that the proposed methodology remains highly reliable for moderately heavy-tailed data, while relegating the treatment of extreme non-Gaussian extensions to future research.

\subsubsection{Scenario A: Case Studies}

In the first simulation scenario, we evaluate the performance of the proposed methods across nine distinct setting through varying $\rho$ and $\alpha$ with $K=1000$. The primary objective is to demonstrate that the proposed MLE can identify the serial dependence and the cross-sectional correlation structure. To provide a comprehensive assessment, Table \ref{tab:scenario_a_combined} reports the estimation means alongside their corresponding  RMSEs in parentheses, allowing for a simultaneous evaluation of overall accuracy and estimation precision. The table provides the MLE against the standard two-step IFM approach, offering a clear comparison of their relative estimation efficiencies.

\begin{table}[htbp]
	\centering
	\caption[Scenario A: Estimation performance across nine parameter settings]{Scenario A: Estimation performance across nine parameter settings ($K=1000$)} 
	\label{tab:scenario_a_combined}
	\renewcommand{\arraystretch}{2.3} % 增加行高，避免括弧與下一行太擠
	\resizebox{\textwidth}{!}{
		\begin{tabular}{cc | cccccc | cccccc}
			\toprule
			\multicolumn{2}{c|}{\textbf{True}} & \multicolumn{6}{c|}{\textbf{MLE}} & \multicolumn{6}{c}{\textbf{IFM}} \\
			$\rho$ & $\alpha$ & $\alpha$ & $\mu_1$ & $\mu_2$ & $\Sigma_{11}$ & $\Sigma_{22}$ & $\Sigma_{12}$ & $\alpha$ & $\mu_1$ & $\mu_2$ & $\Sigma_{11}$ & $\Sigma_{22}$ & $\Sigma_{12}$ \\
			\midrule
			% Row 1
			-0.4 & 0.1 & \makecell{0.1002 \\ (0.0379)} & \makecell{$-0.5570$ \\ (0.2352)} & \makecell{$-0.0980$ \\ (0.1736)} & \makecell{54.9305 \\ (2.4961)} & \makecell{27.9560 \\ (1.2473)} & \makecell{$-15.6649$ \\ (1.3096)} & \makecell{0.0999 \\ (0.0378)} & \makecell{$-0.5573$ \\ (0.2354)} & \makecell{$-0.0976$ \\ (0.1741)} & \makecell{54.9351 \\ (2.5026)} & \makecell{27.9541 \\ (1.2511)} & \makecell{$-15.6676$ \\ (1.3126)} \\
			\midrule
			% Row 2
			-0.4 & 2.0 & \makecell{2.0124 \\ (0.2105)} & \makecell{$-0.5718$ \\ (0.4222)} & \makecell{$-0.1043$ \\ (0.2089)} & \makecell{54.9780 \\ (3.1089)} & \makecell{28.0016 \\ (1.5193)} & \makecell{$-15.6911$ \\ (1.1534)} & \makecell{1.9733 \\ (0.2821)} & \makecell{$-0.5532$ \\ (0.4864)} & \makecell{$-0.0983$ \\ (0.2419)} & \makecell{54.6467 \\ (4.6556)} & \makecell{27.9166 \\ (1.8419)} & \makecell{$-15.7552$ \\ (1.4604)} \\
			\midrule
			% Row 3 (Bold)
			-0.4 & 5.0 & \makecell{\textbf{5.0802} \\ \textbf{(0.7291)}} & \makecell{\textbf{$-0.5973$} \\ \textbf{(0.6517)}} & \makecell{\textbf{$-0.1111$} \\ \textbf{(0.2638)}} & \makecell{\textbf{55.1579} \\ \textbf{(4.1573)}} & \makecell{\textbf{28.0996} \\ \textbf{(2.0770)}} & \makecell{\textbf{$-15.7574$} \\ \textbf{(1.2536)}} & \makecell{4.7568 \\ (1.1979)} & \makecell{$-0.5659$ \\ (1.0041)} & \makecell{$-0.1036$ \\ (0.4123)} & \makecell{53.9803 \\ (8.4441)} & \makecell{27.8609 \\ (3.2700)} & \makecell{$-16.0409$ \\ (1.9059)} \\
			\midrule
			% Row 4
			0.1 & 0.1 & \makecell{0.1000 \\ (0.0379)} & \makecell{$-0.5564$ \\ (0.2354)} & \makecell{$-0.1007$ \\ (0.1790)} & \makecell{54.9496 \\ (2.5023)} & \makecell{27.9571 \\ (1.2687)} & \makecell{3.9342 \\ (1.2123)} & \makecell{0.0997 \\ (0.0377)} & \makecell{$-0.5566$ \\ (0.2355)} & \makecell{$-0.1004$ \\ (0.1793)} & \makecell{54.9535 \\ (2.5096)} & \makecell{27.9533 \\ (1.2698)} & \makecell{3.9323 \\ (1.2174)} \\
			\midrule
			% Row 5
			0.1 & 2.0 & \makecell{2.0124 \\ (0.2106)} & \makecell{$-0.5718$ \\ (0.4222)} & \makecell{$-0.1129$ \\ (0.3315)} & \makecell{54.9782 \\ (3.1090)} & \makecell{28.0038 \\ (1.6952)} & \makecell{3.9225 \\ (0.9618)} & \makecell{1.9733 \\ (0.2821)} & \makecell{$-0.5532$ \\ (0.4864)} & \makecell{$-0.0994$ \\ (0.3886)} & \makecell{54.6467 \\ (4.6556)} & \makecell{27.8023 \\ (2.4853)} & \makecell{3.7265 \\ (2.0741)} \\
			\midrule
			% Row 6 (Bold)
			0.1 & 5.0 & \makecell{\textbf{5.0807} \\ \textbf{(0.7293)}} & \makecell{\textbf{$-0.5978$} \\ \textbf{(0.6519)}} & \makecell{\textbf{$-0.1305$} \\ \textbf{(0.5023)}} & \makecell{\textbf{55.1604} \\ \textbf{(4.1574)}} & \makecell{\textbf{28.0920} \\ \textbf{(2.1580)}} & \makecell{\textbf{3.9194} \\ \textbf{(0.6988)}} & \makecell{4.7553 \\ (1.1972)} & \makecell{$-0.5643$ \\ (1.0020)} & \makecell{$-0.1088$ \\ (0.7965)} & \makecell{53.9633 \\ (8.4242)} & \makecell{27.3921 \\ (4.6011)} & \makecell{3.1542 \\ (3.4221)} \\
			\midrule
			% Row 7
			0.4 & 0.1 & \makecell{0.1000 \\ (0.0380)} & \makecell{$-0.5564$ \\ (0.2351)} & \makecell{$-0.1023$ \\ (0.1803)} & \makecell{54.9509 \\ (2.4918)} & \makecell{27.9659 \\ (1.2697)} & \makecell{15.6918 \\ (1.3185)} & \makecell{0.0997 \\ (0.0379)} & \makecell{$-0.5566$ \\ (0.2353)} & \makecell{$-0.1021$ \\ (0.1804)} & \makecell{54.9553 \\ (2.4981)} & \makecell{27.9623 \\ (1.2705)} & \makecell{15.6912 \\ (1.3242)} \\
			\midrule
			% Row 8
			0.4 & 2.0 & \makecell{2.0124 \\ (0.2105)} & \makecell{$-0.5718$ \\ (0.4222)} & \makecell{$-0.1167$ \\ (0.3829)} & \makecell{54.9785 \\ (3.1090)} & \makecell{28.0016 \\ (1.7789)} & \makecell{15.6908 \\ (1.2867)} & \makecell{1.9733 \\ (0.2821)} & \makecell{$-0.5532$ \\ (0.4864)} & \makecell{$-0.1002$ \\ (0.4487)} & \makecell{54.6467 \\ (4.6556)} & \makecell{27.7352 \\ (2.9362)} & \makecell{15.4374 \\ (2.7051)} \\
			\midrule
			% Row 9 (Bold)
			0.4 & 5.0 & \makecell{\textbf{5.0804} \\ \textbf{(0.7293)}} & \makecell{\textbf{$-0.5974$} \\ \textbf{(0.6518)}} & \makecell{\textbf{$-0.1382$} \\ \textbf{(0.5981)}} & \makecell{\textbf{55.1584} \\ \textbf{(4.1587)}} & \makecell{\textbf{28.0826} \\ \textbf{(2.1914)}} & \makecell{\textbf{15.7272} \\ \textbf{(1.3634)}} & \makecell{4.7568 \\ (1.1979)} & \makecell{$-0.5659$ \\ (1.0041)} & \makecell{$-0.1127$ \\ (0.9479)} & \makecell{53.9803 \\ (8.4441)} & \makecell{27.1362 \\ (5.5163)} & \makecell{14.7713 \\ (4.7889)} \\
			\bottomrule
	\end{tabular}}
	\begin{flushleft}
		\footnotesize \textit{Note:} Values in parentheses represent the RMSEs.
	\end{flushleft}
\end{table}

Table \ref{tab:scenario_a_combined} highlights the superior structural efficiency of the MLE approach. In low-dependence regimes ($\alpha=0.1$), both estimation methods perform equally well in terms of estimation accuracy and precision. However, as the tail dependence intensifies ($\alpha=5.0$), Table \ref{tab:scenario_a_combined} reveals that the IFM estimate exhibits significant underestimation, particularly for the copula parameter $\hat{\alpha}_{IFM} \approx 4.76$. Furthermore, Table \ref{tab:scenario_a_combined} shows a sharp increase in RMSE for the IFM method across all structural parameters, reflecting the accumulation of estimation errors in the multi-stage procedure. In contrast, the MLE remains remarkably robust and  accurate with smaller RMSEs across the entire parameter space. Furthermore, the estimation precision of the marginal parameters ($\mu, \Sigma$) remains consistently stable regardless of the serial dependence intensity $\alpha$. This results provides strong evidence that the hierarchical likelihood optimization successfully decouples the cross-sectional correlation from the latent temporal dynamics, preventing estimation interference even under extreme dependence structures.

\subsubsection{Scenario B: Finite-Sample Convergence}

The second scenario investigates the finite-sample estimation performance across sample sizes. It is fundamental in statistical modeling to ensure that estimation errors systematically diminish as the data volume expands. We evaluate the convergence behavior across increasing sample sizes ($K \in \{100, 250, 500, 750, 1000\}$) under one specific representative structural setting ($\rho=0.4, \alpha=0.1$). 

Table \ref{tab:scenario_b_convergence} reports both the estimation means and the RMSEs for all structural parameters. Comparing the MLE with the IFM approach, the numerical results confirm that as the sample size $K$ increases, the RMSEs for all structural parameters decrease. Furthermore, the mean estimates consistently concentrate around the true parameter values especially in the case of the relatively large sample size leading to the finite-sample accuracy. 

%While both the MLE and IFM estimators exhibit comparable asymptotic behaviors in this low-dependence regime, the systematic reduction in estimation error validates the theoretical soundness of the proposed hierarchical likelihood optimization. 

% --- Table: Scenario B (Convergence) ---
\begin{table}[H]
	\centering
	\renewcommand{\arraystretch}{0.5} % 調整為 0.8 避免過於擁擠
	\setlength{\tabcolsep}{8pt}
	\caption{Scenario B: Finite-sample estimation performance across sample sizes.}
	\label{tab:scenario_b_convergence}
	\resizebox{\textwidth}{!}{
		\begin{tabular}{cc | cc | cc | cc | cc | cc | cc}
			\toprule
			\multirow{2}{*}{\textbf{$K$}} & \multirow{2}{*}{\textbf{Method}} & \multicolumn{2}{c|}{$\alpha$} & \multicolumn{2}{c|}{$\mu_1$} & \multicolumn{2}{c|}{$\mu_2$} & \multicolumn{2}{c|}{$\Sigma_{11}$} & \multicolumn{2}{c|}{$\Sigma_{22}$} & \multicolumn{2}{c}{$\Sigma_{12}$} \\
			& & \textbf{Mean} & \textbf{RMSE} & \textbf{Mean} & \textbf{RMSE} & \textbf{Mean} & \textbf{RMSE} & \textbf{Mean} & \textbf{RMSE} & \textbf{Mean} & \textbf{RMSE} & \textbf{Mean} & \textbf{RMSE} \\
			\midrule
			\multicolumn{2}{c|}{\textbf{True Value}} & \multicolumn{2}{c|}{0.1000} & \multicolumn{2}{c|}{$-0.5500$} & \multicolumn{2}{c|}{$-0.1000$} & \multicolumn{2}{c|}{55.0000} & \multicolumn{2}{c|}{28.0000} & \multicolumn{2}{c}{15.6971} \\
			\midrule
			% --- 新增 N=100 區塊 ---
			\multirow{2}{*}{\textbf{100}} 
			& \textbf{MLE} & 0.1476 & 0.1181 & $-0.5781$ & 0.7724 & $-0.1154$ & 0.5541 & 54.7138 & 7.9040 & 27.9414 & 4.1390 & 15.7871 & 4.2905 \\
			& \textbf{IFM} & 0.1415 & 0.1085 & $-0.5755$ & 0.7717 & $-0.1117$ & 0.5526 & 54.6441 & 7.8911 & 27.8635 & 4.1068 & 15.7233 & 4.2611 \\
			\midrule
			\multirow{2}{*}{\textbf{250}} 
			& \textbf{MLE} & 0.1142 & 0.0692 & $-0.5539$ & 0.4810 & $-0.0977$ & 0.3550 & 55.0258 & 4.9880 & 27.9677 & 2.4959 & 15.7548 & 2.6701 \\
			& \textbf{IFM} & 0.1126 & 0.0676 & $-0.5538$ & 0.4798 & $-0.0970$ & 0.3553 & 55.0224 & 4.9912 & 27.9493 & 2.5020 & 15.7397 & 2.6833 \\
			\midrule
			\multirow{2}{*}{\textbf{500}} 
			& \textbf{MLE} & 0.1035 & 0.0511 & $-0.5528$ & 0.3364 & $-0.0983$ & 0.2533 & 54.9297 & 3.5644 & 27.9394 & 1.7733 & 15.6587 & 1.8885 \\
			& \textbf{IFM} & 0.1029 & 0.0506 & $-0.5525$ & 0.3366 & $-0.0982$ & 0.2533 & 54.9243 & 3.5665 & 27.9341 & 1.7775 & 15.6495 & 1.8944 \\
			\midrule
			\multirow{2}{*}{\textbf{750}} 
			& \textbf{MLE} & 0.1020 & 0.0428 & $-0.5545$ & 0.2709 & $-0.0997$ & 0.2136 & 54.9603 & 2.8826 & 27.9524 & 1.4619 & 15.6687 & 1.5211 \\
			& \textbf{IFM} & 0.1016 & 0.0426 & $-0.5546$ & 0.2713 & $-0.0995$ & 0.2137 & 54.9615 & 2.8866 & 27.9475 & 1.4634 & 15.6655 & 1.5317 \\
			\midrule
			\multirow{2}{*}{\textbf{1000}} 
			& \textbf{MLE} & 0.1000 & 0.0380 & $-0.5564$ & 0.2351 & $-0.1023$ & 0.1803 & 54.9509 & 2.4918 & 27.9659 & 1.2697 & 15.6918 & 1.3185 \\
			& \textbf{IFM} & 0.0997 & 0.0379 & $-0.5566$ & 0.2353 & $-0.1021$ & 0.1804 & 54.9553 & 2.4981 & 27.9623 & 1.2705 & 15.6912 & 1.3242 \\
			\bottomrule
	\end{tabular}}
\end{table}

\subsubsection{Scenario C: Empirical Calibration Validation}
\textcolor{black}{To bridge the theoretical simulations with the specific real-world applications in Section \ref{ES}, we consider the performance of the proposed method relied on the parameters and sample sizes given by two specific periods: the 2008 GFC and the 2020 COVID-19.}
%By configuring the Data Generating Process with representative extreme parameters---characterized by exceptionally high marginal variances and intense lower-tail dependence---and , 
We rigorously pressure-test the estimator to ensure its practical reliability based on the “short sample lengths observed in the actual data ($K=165$ and $K=167$). To capture the core structural severity of the crises without overfitting to finite-sample noise, these true parameter values are systematically rounded from the MLEs of the empirical datasets.  

Table \ref{tab:scenario_c_empirical} shows results under severe empirical conditions. Owing to the smaller serial dependence in these two scenarios, the MLE and two-step IFM approach provide competitive results in terms of the smaller RMSEs. Furthermore, despite the larger volatility (e.g., elevated $\Sigma_{11} \approx 143$) with weaker serial asymmetric dependence, and the constrained finite-sample environments ($K < 170$), corresponding to approximately eight months of trading days), the proposed method provides efficient estimation in terms of mean estimates and the RMSEs.

\begin{table}[H]
	\centering
	\caption[Scenario C: Estimation based on Rounded Empirical Parameters]{Scenario C: Estimation based on Rounded Empirical Parameters}
	\label{tab:scenario_c_empirical}
	\resizebox{\textwidth}{!}{
		\begin{tabular}{l c | cc | cc | cc | cc | cc | cc}
			\toprule
			\multirow{2}{*}{\textbf{Crisis Period}} & \multirow{2}{*}{\textbf{Method}} & \multicolumn{2}{c|}{$\alpha$} & \multicolumn{2}{c|}{$\mu_1$} & \multicolumn{2}{c|}{$\mu_2$} & \multicolumn{2}{c|}{$\Sigma_{11}$} & \multicolumn{2}{c|}{$\Sigma_{22}$} & \multicolumn{2}{c}{$\Sigma_{12}$} \\
			& & \textbf{Mean} & \textbf{RMSE} & \textbf{Mean} & \textbf{RMSE} & \textbf{Mean} & \textbf{RMSE} & \textbf{Mean} & \textbf{RMSE} & \textbf{Mean} & \textbf{RMSE} & \textbf{Mean} & \textbf{RMSE} \\
			\midrule
			\multicolumn{14}{l}{\textbf{2008 GFC  ($K=165$)}} \\
			\midrule
			\multicolumn{2}{c|}{\textbf{True Value}} & \multicolumn{2}{c|}{0.1400} & \multicolumn{2}{c|}{$-1.0000$} & \multicolumn{2}{c|}{0.1600} & \multicolumn{2}{c|}{143.0000} & \multicolumn{2}{c|}{58.0000} & \multicolumn{2}{c}{38.0000} \\
			\midrule
			& \textbf{MLE} & 0.1543 & 0.0918 & $-1.0127$ & 1.0049 & 0.1337 & 0.6580 & 142.4108 & 16.2081 & 57.8016 & 6.4248 & 38.0545 & 7.6362 \\
			& \textbf{IFM} & 0.1509 & 0.0882 & $-1.0141$ & 1.0049 & 0.1361 & 0.6589 & 142.4115 & 16.2994 & 57.7525 & 6.4461 & 38.0489 & 7.6689 \\
			\midrule
			\multicolumn{14}{l}{\textbf{2020 COVID-19 ($K=167$)}} \\
			\midrule
			\multicolumn{2}{c|}{\textbf{True Value}} & \multicolumn{2}{c|}{0.2400} & \multicolumn{2}{c|}{0.1700} & \multicolumn{2}{c|}{0.2500} & \multicolumn{2}{c|}{22.0000} & \multicolumn{2}{c|}{17.0000} & \multicolumn{2}{c}{14.0000} \\
			\midrule
			& \textbf{MLE} & 0.2411 & 0.1143 & 0.1671 & 0.4058 & 0.2424 & 0.3843 & 21.8796 & 2.5439 & 16.8836 & 1.9723 & 13.9285 & 1.9339 \\
			& \textbf{IFM} & 0.2358 & 0.1105 & 0.1661 & 0.4063 & 0.2433 & 0.3850 & 21.8799 & 2.5729 & 16.8750 & 1.9948 & 13.9298 & 1.9670 \\
			\bottomrule
	\end{tabular}}
\end{table}

\subsection*{Summary}
The simulation studies provide a comprehensive validation of the proposed VCBMM framework. The numerical evaluations yield three critical conclusions. First, both the IFM and MLE methods provide to be highly capable estimation in terms of mean estimates and the RMSEs. Moreover, the performance of the MLE demonstrates enhanced structural efficiency, particularly in high-dependence structure where it optimally prevents cross-parameter interference. Second, investigate that the framework exhibits the RMSE for the MLE diminishing as sample sizes increase as we expected. Third, the empirical calibration in Scenario C confirms that the estimates remains highly accurate even under the exact finite-sample constraints and extreme volatility levels observed in historical financial data. Building on these results, we then proceed to apply the framework to empirical data from the 2008 GFC and the 2020 COVID-19 to characterize the lower tail dependence through the Gaussian-Clayton VCBMM. 

%Finally, the misspecification test in Scenario D confirms the estimator's robustness to mild departures from the normality assumption. The Gaussian likelihood yields highly consistent estimates for moderately heavy-tailed data ($\nu \ge 30$), and predictably captures theoretical variance inflation under extreme tail settings, securing its reliability for standard empirical applications.

%In summary, the proposed model is not only theoretically sound but also practically robust for modeling multivariate dependence structures. 

\section{Empirical Studies}\label{ES}

We analyze the daily log returns of bivariate stock portfolios given by 
\[
y_t=100\times\left(\log p_{t}-\log p_{t-1}\right)
\]
where $p_t$ is the vector of adjusted closing prices for the selected assets. The prices were retrieved from Yahoo Finance using the \texttt{quantmod} package \citep{quantmod} within the R programming environment \citep{r_core}. We select Citigroup Inc. (C) and Ford Motor Company (F), defining the event window from September 1, 2008, to April 30, 2009. Conversely, to examine an exogenous real-economy shock, we select the Boeing Company (BA) and Marriott International (MAR) during the COVID-19 pandemic. To ensure regime stability and avoid the initial transitional noise of the outbreak, the COVID-19 window is carefully specified from April 1, 2020, to November 30, 2020. %This event-window approach effectively isolates specific crisis regimes, %avoiding structural breaks caused by mixing tranquil and volatile periods. 

To rigorously benchmark the proposed VCBMM, we select two standard alternatives for comparison: the independent multivariate Gaussian (IMG) model without serial dependence, and the vector autoregressive model of order one (VAR(1)) model \citep{sims1980macroeconomics, lutkepohl2005new,tsay2013multivariate}, which represents the traditional econometric approach to capturing linear temporal dynamics.

%%%%%%%%%%%%%%%%%%%%%%%%%%%%%%%%%%%%%%%%%%%%%%%%%%%%%%%%%%%%%%%%%%%%%%%%%%%
%%%%%%%%%%%%%%%%%%%%%%%%%%%%%% 2008 %%%%%%%%%%%%%%%%%%%%%%%%%%%%%%%%%%%%%%%
%%%%%%%%%%%%%%%%%%%%%%%%%%%%%%%%%%%%%%%%%%%%%%%%%%%%%%%%%%%%%%%%%%%%%%%%%%%
\subsection{Empirical Analysis I: 2008 GFC}\label{2008}
We first evaluate the model performance during the 2008 GFC. The estimation for the proposed VCBMM is provided in Table \ref{tab:params_2008}. In addition, the estimation for the IMG and VAR(1) models is shown in Appendix \ref{GFC2008}.
Table \ref{tab:params_2008} reports the MLEs for the 2008 GFC. To provide a robust assessment of parameter uncertainty, the corresponding 95\% confidence intervals are constructed using a parametric bootstrap resampling procedure. This parametric approach is explicitly chosen to preserve the latent Markovian structure inherent in the time series, which would otherwise be destroyed by standard non-parametric resampling.

\begin{figure}[H]
	\centering
	\includegraphics[width=0.85\textwidth]{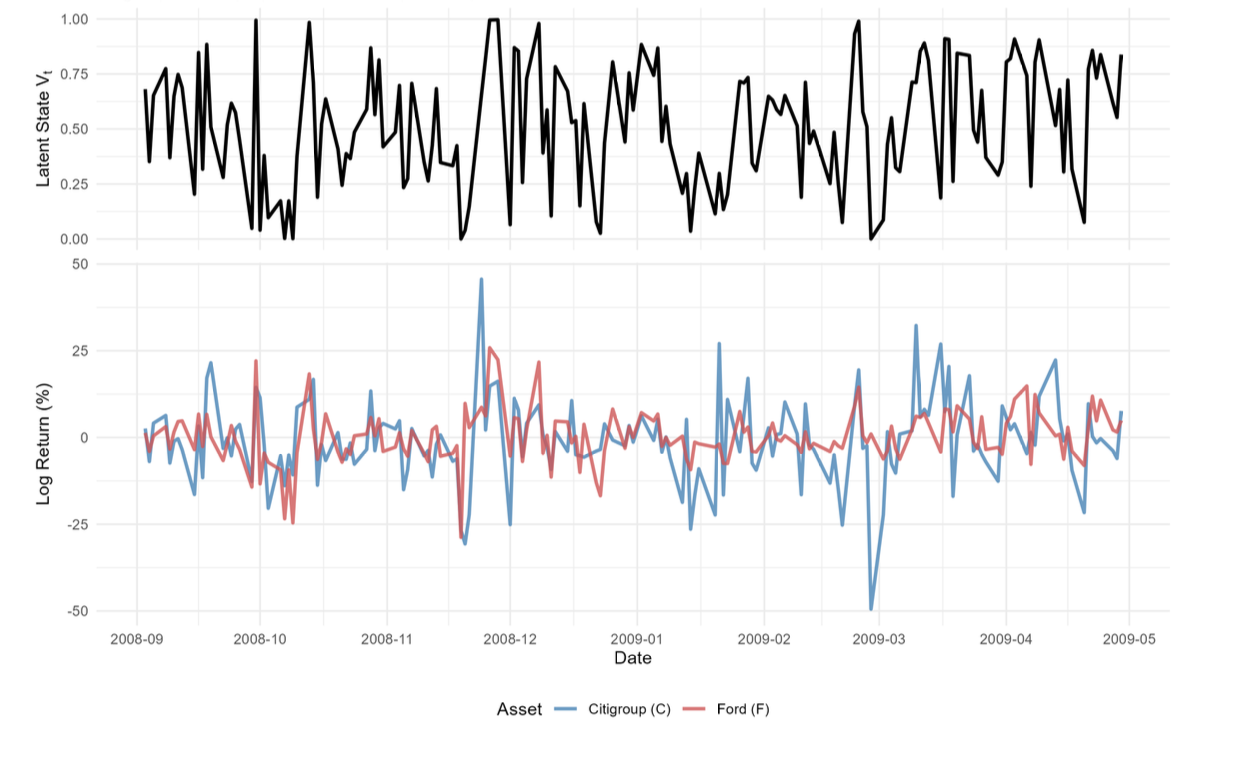} 
	\caption[Comparison of latent signal and returns (2008 GFC)]{Fitted aggregate variables (top) and daily log returns (bottom) for the Citigroup–Ford sample, September 2008–April 2009.} 
	%The extracted latent state captures the joint distress underlying the severe realized market drops (bottom).}
	\label{fig:latent_2008}
\end{figure}

Citigroup’s estimated marginal variance is 142.7291. The estimated Clayton parameter is
$\hat{\alpha} = 0.1419$, with a reported 95\% parametric bootstrap interval of $[0.0147,0.3672]$. Under
the fitted copula, this corresponds to a lower-tail dependence coefficient of approximately
0.0076 between consecutive aggregate variables. Figure \ref{fig:latent_2008} displays the fitted aggregate
variables and bivariate log returns.
%
%First, observe that the covariance matrix captures the extreme instability of the period, with Citigroup's variance reaching an exceptional 142.73. Second, and most critically, the estimates for the Clayton parameter $\hat{\alpha} = 0.1419$ is statistically significant, as the corresponding 95\% bootstrap confidence interval excludes zero given by $[0.0147, 0.3672]$. 
%\textcolor{black}{This provides statistical evidence of lower-tail dependence confirming that there exists contagion channel of the banking and manufacturing sectors during the liquidity crisis.} Figure \ref{fig:latent_2008} visualizes the extracted latent aggregate variable $V_t$ alongside the observed log returns. 

Table 5 compares the three models using the log-likelihood, AIC, and BIC. The VCBMM has the
lowest BIC, whereas the VAR(1) has the lowest AIC. The BIC differences favoring the VCBMM
are approximately 0.65 relative to the IMG model and 0.77 relative to the VAR(1).

%Table \ref{tab:comp_2008} reports the goodness-of-fit metrics across these models. The results highlight a critical trade-off between model fit and parsimony. \textcolor{black}{Compared to the VAR(1) model with the higher parameterized linear lag structure ($k=9$) , observe that the Bayesian Information Criterion (BIC), which strictly penalizes model complexity, the proposed VCBMM ($k=6$) outperforms VAR(1) benchmark. 
%This demonstrates that for high-volatility event windows, the parsimonious VCBMM framework exhibits higher structural efficiency in capturing the lower-tail risk for consecutive aggregate variables $V_t$ and $V_{t-1}$ where indicates financial contagion in serial dependence.}

%While the VAR(1) model achieves a larger log-likelihood due to its highly parameterized linear lag structure ($k=9$) and a lower Akaike Information Criterion (AIC). 
%%which traditionally excels in conditional mean forecasting---it suffers from potential overfitting in finite extreme regimes. }
%When evaluated under  

\begin{table}[H]
	\centering
	\renewcommand{\arraystretch}{0.5} % 稍微收緊行高
	\setlength{\tabcolsep}{8pt}      % 調整欄位間距，避免表格過寬或過窄
	\caption[VCBMM Parameter Estimates (2008 GFC)]{VCBMM Parameter Estimates and 95\% Parametric Bootstrap Confidence Intervals: 2008 GFC}
	\label{tab:params_2008}
	\begin{tabular}{lcc}
		\toprule
		\textbf{Parameter} & \textbf{Estimate (MLE)} & \textbf{95\% Bootstrap CI [Lower, Upper]} \\
		\midrule
		\multicolumn{3}{l}{\textit{Marginal Means}} \\
		$\mu_{Citi}$ & -0.9985 & [-2.9948, 0.8910] \\
		$\mu_{Ford}$ & 0.1570 & [-1.1463, 1.3841] \\
		\midrule
		\multicolumn{3}{l}{\textit{Covariance Structure}} \\
		$\Sigma_{11}$ (Citigroup Variance) & 142.7291 & [112.7742, 176.1804] \\
		$\Sigma_{22}$ (Ford Variance) & 58.0694 & [45.8127, 70.7672] \\
		$\Sigma_{12}$ (Covariance) & 37.5216 & [23.7248, 53.9075] \\
		\midrule
		\multicolumn{3}{l}{\textit{Latent serial dependence}} \\
		$\alpha$ (Clayton Copula) & 0.1419 & [0.0147, 0.3672] \\
		\bottomrule
	\end{tabular}
\end{table}

\begin{table}[H]
	\centering
	\renewcommand{\arraystretch}{0.5} % 稍微收緊行高
	\setlength{\tabcolsep}{8pt}      % 調整欄位間距，避免表格過寬或過窄
	\caption{Model Selection Results: 2008 GFC (Citigroup/Ford)}
	\label{tab:comp_2008}
	\begin{tabular}{lcccc}
		\toprule
		\textbf{Model} & \textbf{Log-Likelihood} & \textbf{Parameters ($k$)} & \textbf{AIC} & \textbf{BIC} \\
		\midrule
		IMG        & -1198.6587 & 5 & 2407.3175 & 2422.8472 \\
		VAR(1)         & \textbf{-1188.5077} & 9 & \textbf{2395.0153} & 2422.9688 \\
		VCBMM (Proposed) & -1195.7797 & \textbf{6} & 2403.5594 & \textbf{2422.1950} \\
		\bottomrule
	\end{tabular}
\end{table}

\subsection{Empirical Analysis II: 2020 COVID-19}\label{2020}

To assess the generalizability of the VCBMM, we further apply the proposed method to the data in 2020 COVID-19 which is an exogenous real-economy shock fundamentally different from the 2008 GFC. The estimation results for the proposed VCBMM are provided in Table \ref{tab:params_2020}. In addition, the estimates for the baseline IMG and VAR(1) models are reported in Appendix \ref{COVID2019}. 

%Similar to the previous analysis, we first present the MLEs to characterize the structural dependence of the pandemic period before proceeding to the model selection results. 

Table \ref{tab:params_2020} reports the fitted parameters and parametric bootstrap intervals for the Boeing–
Marriott sample. The estimated Clayton parameter is $\hat{\alpha} = 0.2363$, with a reported
95\% parametric bootstrap interval of [0.0465,0.4690]. The fitted lower-tail dependence
coefficient between consecutive aggregate variables is approximately 0.0532. Figure 3
displays the fitted aggregate variables and bivariate log returns. Table 7 shows that the VCBMM has the lowest BIC. Nevertheless, the VAR(1) has the lowest AIC. Observe that the BIC
differences favoring VCBMM are approximately 0.75 relative to the IMG model
and 4.61 relative to the VAR(1).

\begin{figure}[H]
	\centering
	\includegraphics[width=0.85\textwidth]{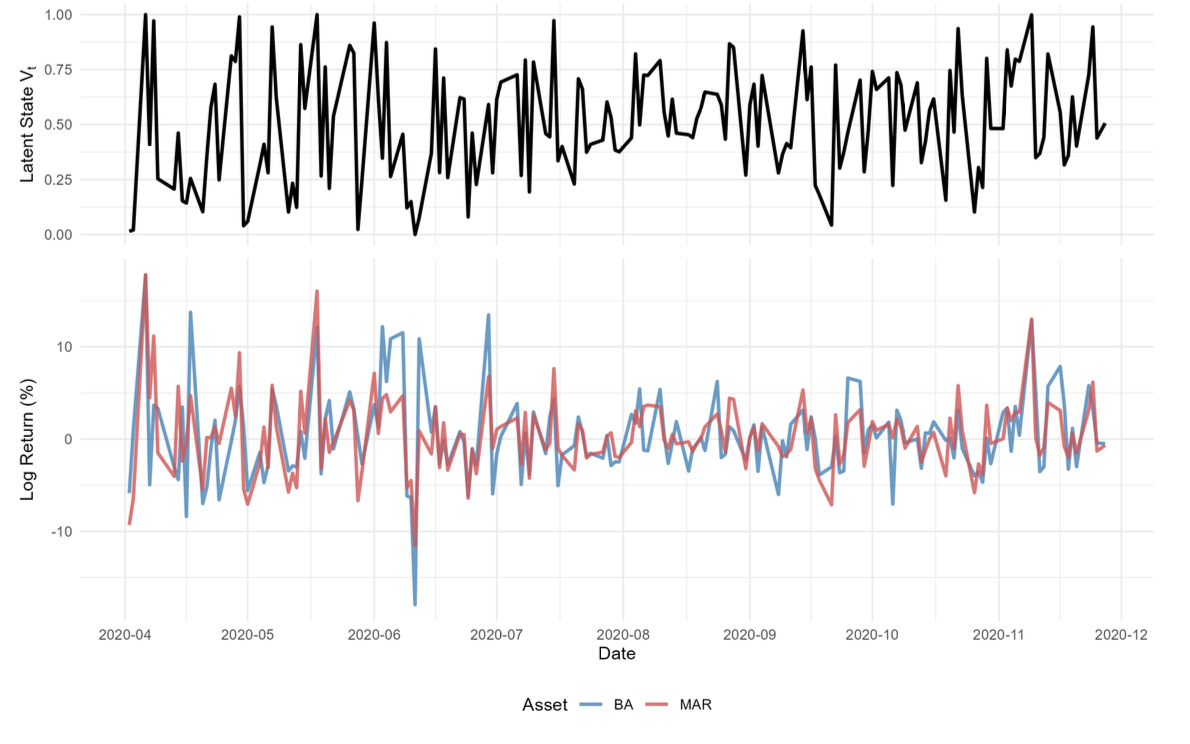} 
	\caption[Comparison of Latent Signal and Market Returns (2020 COVID-19)]{Fitted aggregate variables (top) and daily log returns (bottom) for the Boeing–Marriott
sample, April–November 2020.}
	\label{fig:latent_2020}
\end{figure}

\begin{table}[H]
	\centering
	\renewcommand{\arraystretch}{0.5}
	\setlength{\tabcolsep}{8pt}
	\caption[VCBMM Parameter Estimates (2020 COVID-19)]{VCBMM Parameter Estimates and 95\% Parametric Bootstrap Confidence Intervals: 2020 COVID-19}
	\label{tab:params_2020}
	\begin{tabular}{lcc}
		\toprule
		\textbf{Parameter} & \textbf{Estimate (MLE)} & \textbf{95\% Bootstrap CI [Lower, Upper]} \\
		\midrule
		\multicolumn{3}{l}{\textit{Marginal Means}} \\
		$\mu_{Boeing}$ & 0.1745 & [-0.6395, 0.9512] \\
		$\mu_{Marriott}$ & 0.2499 & [-0.5312, 0.9847] \\
		\midrule
		\multicolumn{3}{l}{\textit{Covariance Structure}} \\
		$\Sigma_{11}$ (Boeing Variance) & 21.9368 & [17.2219, 27.1635] \\
		$\Sigma_{22}$ (Marriott Variance) & 16.8028 & [13.1043, 20.7255] \\
		$\Sigma_{12}$ (Covariance) & 13.9412 & [10.3439, 18.0393] \\
		\midrule
		\multicolumn{3}{l}{\textit{Latent serial dependence}} \\
		$\alpha$ (Clayton Copula) & 0.2363 & [0.0465, 0.4690] \\
		\bottomrule
	\end{tabular}
\end{table}

\begin{table}[H]
	\centering
	\renewcommand{\arraystretch}{0.5}
	\setlength{\tabcolsep}{5pt}
	\caption{Model Selection Results: 2020 COVID-19 (Boeing/Marriott)}
	\label{tab:comp_2020}
	\begin{tabular}{lcccc}
		\toprule
		\textbf{Model} & \textbf{Log-Likelihood} & \textbf{Parameters ($k$)} & \textbf{AIC} & \textbf{BIC} \\
		\midrule
		IMG & -903.4325 & 5 & 1816.8650 & 1832.4550 \\
		VAR(1) & \textbf{-895.1253} & 9 & \textbf{1808.2506} & 1836.3126 \\
		VCBMM (Proposed) & -900.4967 & \textbf{6} & 1812.9934 & \textbf{1831.7014} \\
		\bottomrule
	\end{tabular}
\end{table}

\subsection*{Summary}
The two applications illustrate estimation of the Gaussian–Clayton specification for bivariate daily log returns. The fitted copula describes lower-tail serial dependence between aggregate variables. The proposed VCBMM has the lowest BIC in the 2008 GFC and the 2020 COVID-19, whereas VAR(1) has the lowest AIC. The results show that the proposed model is the competitive one compared to the two benchmarks. Furthermore, the weak lower tail serial dependence can be described through $\hat\alpha>0$ in both scenarios based on the analysis of the Gaussian–Clayton VCBMM. The obtained serial dependence structure could be treated as an alert in real application.

\section{Conclusions}\label{CON}

We propose the VCBMM in which serial dependence is mediated by scalar aggregate variables. For the Gaussian–Clayton specification, we provide an explicit aggregation rule and the corresponding likelihood-based joint estimation procedure. The
simulations shows that the MLE generally have smaller RMSEs under stronger dependence in the
investigated settings. Applications to two bivariate log returns illustrate estimation of
aggregate serial dependence and comparison with the IMG and the VAR(1) benchmarks. 

There are several extension for further study. First, integrating formal change-point detection mechanisms into the VCBMM framework represents a highly valuable extension. The observation windows for crisis regimes were heuristically defined based on historical events. Future research could develop statistical tests---such as likelihood-ratio or residual-based change-point analysis---tailored specifically for vector copula structures to identify structural breaks endogenously. This would enable a more scientific determination of homogeneous market regimes, ensuring that the analyzed windows are purely representative of a specific state before or after a transition.

Second, exploring alternative copula specifications within the established framework presents promising directions for future research. In this study, the empirical application utilized the Gaussian-Clayton structure to specifically capture symmetric spatial correlation and asymmetric lower-tail serial dependence. In the future, we will investigate the implementation of different copula models to accommodate various stylized facts across different disciplines. A Gumbel nesting copula could represent upper-tail dependence between consecutive aggregate variables. This flexibility in various copula families would allow the model to adapt to an even wider range of multivariate time series dynamics beyond financial applications.

%Similarly, applying the Student-$t$ copula for the internal spatial structure could better account for joint heavy-tailed behavior among multivariate sensor networks or physical variables.

Finally, the distributional assumptions can be further broadened. Given the prevalent heavy-tailed and skewed nature of financial assets, it is natural to consider the multivariate Student-t or skew-t marginal distributions in future extensions. The extension would allow the model to identify the heavy tails structure, further expanding the practical impact and predictive power in risk management under extreme market events.

\section*{Acknowledgements}

Li-Hsien Sun receives support from the National Science and Technology Council, Taiwan, under Grant No. NSTC 115-2123-M-035-001 and NSTC 115-2118-M-008 -003 -MY2.

\section*{Data Availability}

The financial time series analyzed in this study are publicly available from Yahoo Finance at \url{https://finance.yahoo.com}. The code is publicly accessible at \url{https://github.com/tpsun7246/VCBMM.git}.

\section*{Declaration of Generative AI Use}

During the preparation of this work, the authors used ChatGPT, developed by OpenAI, solely to support English-language editing and improve the clarity and readability of the manuscript. The authors critically reviewed and revised all AI-generated output and take full responsibility for the content of the published article.

\begin{appendix}

\renewcommand{\thesection}{\Alph{section}}          % 章節顯示為 A, B
\renewcommand{\thesubsection}{\thesection.\arabic{subsection}} % 子章節顯示為 A.1, A.2 (修正 S.1 的問題)
\renewcommand{\thetable}{\thesection.\arabic{table}}       % 表格顯示為 Table A.1
\renewcommand{\theequation}{\thesection.\arabic{equation}} % 公式顯示為 Eq A.1
\setcounter{table}{0}     % 重置表格計數
\setcounter{equation}{0}  % 重置公式計數

\section{Baseline Model Estimates for the 2008 GFC}\label{GFC2008}

Table \ref{tab:iid_mn_2008} and Table \ref{tab:var1_2008} report the parameter estimates for the two baseline models: the IMG and the VAR(1) models, evaluated during the 2008 GFC.

% --- Table 1: iid MN ---
\begin{table}[H]
	\centering
	\small % 縮小字體
	\renewcommand{\arraystretch}{0.9} % 稍微收緊行高
	\setlength{\tabcolsep}{10pt}
	\caption{Parameter Estimates for IMG Model: 2008 GFC}
	\label{tab:iid_mn_2008}
	\begin{tabular}{lc}
		\toprule
		\textbf{Parameter} & \textbf{Estimate (MLE)} \\
		\midrule
		\multicolumn{2}{l}{\textit{Marginal Means}} \\
		$\mu_{Citi}$ & $-1.0894$ \\
		$\mu_{Ford}$ & 0.1147 \\
		\addlinespace
		\multicolumn{2}{l}{\textit{Covariance Structure}} \\
		$\Sigma_{11}$ (Citigroup Variance) & 144.9740 \\
		$\Sigma_{22}$ (Ford Variance) & 58.4401 \\
		$\Sigma_{12}$ (Covariance) & 38.3906 \\
		\bottomrule
	\end{tabular}
\end{table}

% --- Table 2: VAR(1) ---
\begin{table}[H]
	\centering
	\small % 縮小字體
	\renewcommand{\arraystretch}{0.9}
	\setlength{\tabcolsep}{10pt}
	\caption{Parameter Estimates for VAR(1) Model: 2008 GFC}
	\label{tab:var1_2008}
	\begin{tabular}{lc}
		\toprule
		\textbf{Parameter} & \textbf{Estimate (MLE)} \\
		\midrule
		\multicolumn{2}{l}{\textit{Mean Equations (Conditional)}} \\
		$\Phi_{0, Citi}$ (Citigroup Constant) & $-1.1129$ \\
		$\Phi_{11}$ (Lag 1 Citi $\rightarrow$ Citi) & 0.0205 \\
		$\Phi_{12}$ (Lag 1 Ford $\rightarrow$ Citi) & 0.2859 \\
		\addlinespace
		$\Phi_{0, Ford}$ (Ford Constant) & 0.1243 \\
		$\Phi_{21}$ (Lag 1 Citi $\rightarrow$ Ford) & 0.0257 \\
		$\Phi_{22}$ (Lag 1 Ford $\rightarrow$ Ford) & 0.1454 \\
		\addlinespace
		\multicolumn{2}{l}{\textit{Residual Covariance Structure}} \\
		$\Sigma_{11}$ (Citigroup Res. Variance) & 143.0883 \\
		$\Sigma_{22}$ (Ford Res. Variance) & 58.2288 \\
		$\Sigma_{12}$ (Residual Covariance) & 36.3503 \\
		\bottomrule
	\end{tabular}
\end{table}

\section{Baseline Model Estimates for the 2020 COVID-19 }\label{COVID2019}

Table \ref{tab:iid_mn_2020} and Table \ref{tab:var1_2020} report the parameter estimates for the two baseline models: the  IMG and the VAR(1) models, evaluated during the 2020 COVID-19.

% --- Table 1: iid MN (COVID-19) ---
\begin{table}[H]
	\centering
	\small % 縮小字體
	\renewcommand{\arraystretch}{0.9}
	\setlength{\tabcolsep}{10pt}
	\caption{Parameter Estimates for IMG Model: 2020 COVID-19}
	\label{tab:iid_mn_2020}
	\begin{tabular}{lc}
		\toprule
		\textbf{Parameter} & \textbf{Estimate (MLE)} \\
		\midrule
		\multicolumn{2}{l}{\textit{Marginal Means}} \\
		$\mu_{Boeing}$ & 0.3022 \\
		$\mu_{Marriott}$ & 0.3703 \\
		\addlinespace
		\multicolumn{2}{l}{\textit{Covariance Structure}} \\
		$\Sigma_{11}$ (Boeing Variance) & 21.4127 \\
		$\Sigma_{22}$ (Marriott Variance) & 16.1921 \\
		$\Sigma_{12}$ (Covariance) & 13.2420 \\
		\bottomrule
	\end{tabular}
\end{table}

% --- Table 2: VAR(1) (COVID-19) ---
\begin{table}[H]
	\centering
	\small % 縮小字體
	\renewcommand{\arraystretch}{0.9}
	\setlength{\tabcolsep}{10pt}
	\caption{Parameter Estimates for VAR(1) Model: 2020 COVID-19}
	\label{tab:var1_2020}
	\begin{tabular}{lc}
		\toprule
		\textbf{Parameter} & \textbf{Estimate (MLE)} \\
		\midrule
		\multicolumn{2}{l}{\textit{Mean Equations (Conditional)}} \\
		$\Phi_{0, Boeing}$ (Boeing Constant) & 0.3396 \\
		$\Phi_{11}$ (Lag 1 Boeing $\rightarrow$ Boeing) & $-0.1275$ \\ % 數學模式負號
		$\Phi_{12}$ (Lag 1 Marriott $\rightarrow$ Boeing) & 0.1031 \\
		\addlinespace
		$\Phi_{0, Marriott}$ (Marriott Constant) & 0.4194 \\
		$\Phi_{21}$ (Lag 1 Boeing $\rightarrow$ Marriott) & $-0.0694$ \\ % 數學模式負號
		$\Phi_{22}$ (Lag 1 Marriott $\rightarrow$ Marriott) & 0.0813 \\
		\addlinespace
		\multicolumn{2}{l}{\textit{Residual Covariance Structure}} \\
		$\Sigma_{11}$ (Boeing Res. Variance) & 21.5282 \\
		$\Sigma_{22}$ (Marriott Res. Variance) & 15.9484 \\
		$\Sigma_{12}$ (Residual Covariance) & 13.1037 \\
		\bottomrule
	\end{tabular}
\end{table}

% ========================================================================
% Appendix A: Mathematical Derivations
% ========================================================================
\section{Mathematical Derivations of the Score and Hessian}\label{MLE}
\label{app:derivations}

This appendix provides the detailed mathematical derivation of the gradient (Score vector) and the Hessian matrix for the proposed Gaussian-Clayton VCBMM. These analytical expressions are critical for implementing the Full Maximum Likelihood Estimation (MLE) procedure using Newton-type optimization methods. The derivations rely extensively on standard results from matrix differential calculus and matrix algebra. The notation and differential techniques adopted below follow the frameworks of  \cite{harville1997} and  \cite{magnus1999}.

% 【請在此處貼上您原本的微分推導 LaTeX 代碼】
% Paste your supplementary material latex code here...

\subsection{Log-Likelihood Function}

Let $y_i\in\R^d$, $i=1,\ldots,K$, denote the observed $d$-dimensional data vectors. The marginal distribution of each observation is assumed to be the multivariate Gaussian distribution with mean vector $\mu$ and covariance matrix
\[
\Sigma=LL^{\T},
\]
where $L$ is the corresponding Cholesky factor. Serial dependence between consecutive observations is characterized by a Clayton copula with parameter $\alpha$.

Under this specification, the complete log-likelihood can be decomposed into the Gaussian marginal contribution $L_m$ and the copula contribution $L_c$:
\begin{align}
\nonumber \cL(\mu,L,\alpha)
&=\sum_{i=1}^{K}\log\!\left[
\frac{1}{(2\pi)^{d/2}|LL^{\T}|^{1/2}}
\exp\!\left\{-\frac12(y_i-\mu)^{\T}(LL^{\T})^{-1}(y_i-\mu)\right\}
\right] \\
\nonumber &\quad+\sum_{i=1}^{K-1}\log\!\left[
(1+\alpha)V_i^{-(1+\alpha)}V_{i+1}^{-(1+\alpha)}
\bigl(V_i^{-\alpha}+V_{i+1}^{-\alpha}-1\bigr)^{-(2+1/\alpha)}
\right] \\
&=L_m+L_c,
\end{align}
with  
\begin{equation*}
	\mathcal{L}_m =   -\frac{Kd}{2}\ln(2\pi) - \frac{K}{2}\ln|{{LL^{\T}}}| -  \sum_{i=1}^{K} \frac{1}{2} {z}_i^\top  {z}_i,
\end{equation*}
\begin{equation*}
	\mathcal{L}_c = \sum_{i=1}^{K-1}\Bigg(\ln(1+\alpha) - (1+\alpha)\ln({V}_i {V}_{i+1}) - \left(\frac{1}{\alpha}+2\right)\ln(\mathcal{C}_i) \Bigg) \cdot \boldsymbol{I}(\mathcal{C}_i > 0),
\end{equation*}

\noindent where the intermediate variables are defined as:
\begin{align*}
	{z}_i &=  {L}^{-1}( {y}_i -  {\mu}) ~, \\
	 {R}_i &= \sum_{j=1}^{d} \ln \Phi(( {z}_i)_j) ~, \\
	 {V}_i &= F_{ {R}_i}({R}_i)~, \\
	 {C}_i &=  {V}_i^{-\alpha}+ {V}_{i+1}^{-\alpha}-1, 
\end{align*}
for $i=1,\ldots,K-1.$ Note that the Clayton-copula term is understood on the support
\[
V_i^{-\alpha}+V_{i+1}^{-\alpha}-1>0.
\]
%For later use, define
%\[
%C_i=V_i^{-\alpha}+V_{i+1}^{-\alpha}-1,
%\qquad i=1,\ldots,K-1.
%\]
%

\subsection{Derivatives with respect to Copula Parameter \texorpdfstring{$\alpha$}{alpha}}

Since the Gaussian marginal term does not depend on $\alpha$, all derivatives with respect to $\alpha$ arise from $L_c$. For one adjacent pair $(V_i,V_{i+1})$, the copula log-density can be written as
\[
\ell_{c,i}=\log(1+\alpha)-(1+\alpha)\log(V_iV_{i+1})-
\left(2+\frac1\alpha\right)\log C_i.
\]
Hence the score with respect to $\alpha$ is
\begin{align}
\frac{\partial L_c}{\partial\alpha}
=\sum_{i=1}^{K-1}\Bigg[
&\frac{1}{1+\alpha}-\log(V_iV_{i+1})
+\frac{1}{\alpha^2}\log C_i -\left(2+\frac1\alpha\right)
\frac{\partial_\alpha C_i}{C_i}
\Bigg],
\end{align}
where
\[
\partial_\alpha C_i
=-V_i^{-\alpha}\log V_i-V_{i+1}^{-\alpha}\log V_{i+1}.
\]
The second derivative is therefore
\begin{align}
H_{\alpha\alpha}
=\frac{\partial^2L_c}{\partial\alpha^2}
=\sum_{i=1}^{K-1}\Bigg[
&-\frac{1}{(1+\alpha)^2}
-\frac{2}{\alpha^3}\log C_i
+\frac{2}{\alpha^2}\frac{\partial_\alpha C_i}{C_i} -\left(2+\frac1\alpha\right)
\left[\frac{\partial_{\alpha\alpha}^2C_i}{C_i}
-\left(\frac{\partial_\alpha C_i}{C_i}\right)^2\right]
\Bigg],
\end{align}
with
\[
\partial_{\alpha\alpha}^2C_i
=V_i^{-\alpha}(\log V_i)^2+V_{i+1}^{-\alpha}(\log V_{i+1})^2.
\]
Here and throughout the appendix, the dependence of $V_i$ on the remaining model parameters is kept implicit whenever the differentiation variable is unambiguous.

% \subsection{Derivatives with respect to Mean Vector $\boldsymbol{\mu}$}

\subsection{Derivatives with respect to Mean Vector \texorpdfstring{$ {\mu}$}{mu}}
When derivatives are taken with respect to the mean vector $\boldsymbol{\mu}$, both the Gaussian marginal component and the copula component contribute.

\subsubsection{Gradient Vector}
The score with respect to $\mu$ is
\[
\nabla_\mu\cL=\nabla_\mu L_m+\nabla_\mu L_c,
\]
where
\[
\nabla_\mu L_m=(LL^{\T})^{-1}\sum_{i=1}^{K}(y_i-\mu),
\]
and
\begin{align}
\nabla_\mu L_c
=\sum_{i=1}^{K-1}\Bigg[
&-(1+\alpha)\frac{1}{V_i}\nabla_\mu V_i
-(1+\alpha)\frac{1}{V_{i+1}}\nabla_\mu V_{i+1} -\left(\frac1\alpha+2\right)\frac1{C_i}\nabla_\mu C_i
\Bigg].
\end{align}
The required intermediate derivatives are
\[
\nabla_\mu C_i
=-\alpha V_i^{-\alpha-1}\nabla_\mu V_i
-\alpha V_{i+1}^{-\alpha-1}\nabla_\mu V_{i+1},
\]
\[
\nabla_\mu V_i=f_{R_i}(R_i)\nabla_\mu R_i,
\qquad
\nabla_\mu R_i=-L^{-\T}m(z_i).
\]
Here $m(z_i)$ denotes the vector of inverse Mills ratios, whose $j$th component is
\[
[m(z_i)]_j=\frac{\phi(z_{ij})}{\Phi(z_{ij})}.
\]

\subsubsection{Hessian Matrix}
The Hessian block is
\[
H_{\mu\mu}=\nabla_\mu^2\cL
=-K(LL^{\T})^{-1}
+\sum_{i=1}^{K-1}\left(H_{V_i}^{\mathrm{term}}+H_{V_{i+1}}^{\mathrm{term}}+H_{C_i}^{\mathrm{term}}\right),
\]
where, for $k\in\{i,i+1\}$,
\[
H_{V_k}^{\mathrm{term}}
=-(1+\alpha)\left[-V_k^{-2}(\nabla_\mu V_k)(\nabla_\mu V_k)^{\T}
+V_k^{-1}\nabla_\mu^2V_k\right],
\]
and
\[
H_{C_i}^{\mathrm{term}}
=-\left(\frac1\alpha+2\right)
\left[-C_i^{-2}(\nabla_\mu C_i)(\nabla_\mu C_i)^{\T}
+C_i^{-1}\nabla_\mu^2C_i\right].
\]
The second-order derivatives of the intermediate quantities satisfy
\[
\nabla_\mu^2V_i
=(\nabla_\mu R_i)(\nabla_\mu f_{R_i})^{\T}
+f_{R_i}(R_i)\nabla_\mu^2R_i,
\]
\[
\nabla_\mu f_{R_i}
=f_{R_i}(R_i)\left(1+\frac{d-1}{R_i}\right)\nabla_\mu R_i,
\qquad
\nabla_\mu^2R_i=L^{-\T}{\mathrm{diag}}\bigl(m'(z_i)\bigr)L^{-1}.
\]

\subsection{Derivatives with respect to Cholesky Factor ${L}$}

Let
\[
z_i=L^{-1}(y_i-\mu).
\]
For fixed $\mu$, the differential of $z_i$ with respect to $L$ is
\[
dz_i=-L^{-1}(dL)z_i.
\]

\subsubsection{Gradient Matrix}
For the Gaussian marginal component,
\[
L_m=-K\log|L|-\frac12\sum_{i=1}^{K}z_i^{\T}z_i+\text{constant}.
\]
Using $d\log|L|=\tr(L^{-1}dL)$ and the expression for $dz_i$, one obtains
\[
\nabla_LL_m
=L^{-\T}\left(\sum_{i=1}^{K}z_i z_i^{\T}-KI_d\right)
=L^{-\T}\Omega_m,
\]
where
\[
\Omega_m=\sum_{i=1}^{K}z_i z_i^{\T}-KI_d.
\]

For the copula component, define
\[
\delta_i=\nabla_{z_i}L_c.
\]
Then
\begin{align}
dL_c
&=\sum_{i=1}^{K}\delta_i^{\T}dz_i
=\sum_{i=1}^{K}\delta_i^{\T}\left[-L^{-1}(dL)z_i\right] \\
&=\tr\left[
\left(-L^{-\T}\sum_{i=1}^{K}\delta_i z_i^{\T}\right)^{\T}dL
\right].
\end{align}
Therefore
\[
\nabla_LL_c=-L^{-\T}\sum_{i=1}^{K}\delta_i z_i^{\T}
=-L^{-\T}\Omega_c,
\qquad
\Omega_c=\sum_{i=1}^{K}\delta_i z_i^{\T}.
\]
Consequently, the total gradient with respect to $L$ is
\[
G=\nabla_L\cL=L^{-\T}(\Omega_m-\Omega_c).
\]

\subsubsection{Hessian Matrix}
Let
\[
M=\Omega_m-\Omega_c.
\]
Differentiating $G=L^{-\T}M$ gives
\[
dG=-L^{-\T}(dL)^{\T}L^{-\T}M+L^{-\T}(d\Omega_m-d\Omega_c).
\]
The first term arises from differentiating $L^{-\T}$, whereas the second term is generated by the differential of $M$.

Using the standard vectorization identity
\[
\vecop(AXB)=(B^{\T}\otimes A)\vecop(X)
\]
and the commutation matrix $K_{dd}$, vectorization of the first term yields
\[
T_1=-\bigl(M^{\T}L^{-1}\otimes L^{-\T}\bigr)K_{dd}.
\]
For the second term, define
\[
Q_i=z_i-\delta_i.
\]
Then
\[
dM=\sum_{i=1}^{K}\left[(dz_i-d\delta_i)z_i^{\T}+Q_i(dz_i)^{\T}\right].
\]
Using
\[
dz_i=-L^{-1}(dL)z_i,
\qquad
d\delta_i=\sum_jJ_{ij}\,dz_j,
\]
we identify the remaining Hessian contributions.

The term arising from $Q_i(dz_i)^{\T}$ is
\[
T_2=-\sum_{i=1}^{K}
\left[L^{-1}\otimes\left(L^{-\T}Q_i z_i^{\T}\right)\right]K_{dd}.
\]
The diagonal contribution involving $J_{ii}$ is
\[
T_3=-\sum_{i=1}^{K}
\left[(z_i z_i^{\T})\otimes\left(L^{-\T}(I_d-J_{ii})L^{-1}\right)\right].
\]
The off-diagonal contribution involving $J_{ij}$, $i\neq j$, is
\[
T_{\mathrm{cross}}
=\sum_{i\neq j}
\left[(z_i z_j^{\T})\otimes\left(L^{-\T}J_{ij}L^{-1}\right)\right].
\]
Thus
\[
\boxed{H_{LL}=T_1+T_2+T_3+T_{\mathrm{cross}}.}
\]

% \subsection{Cross-Derivatives: Copula Parameter $\alpha$ and Mean Vector $\boldsymbol{\mu}$}

\subsection[Cross-Derivatives: Copula Parameter alpha and Mean Vector mu]{Cross-Derivatives: Copula Parameter $\alpha$ and Mean Vector $ \mu$}

The mixed derivative involving $\alpha$ and $\mu$ is obtained by differentiating the score $\partial L_c/\partial\alpha$ with respect to $\mu$. Since $L_m$ does not depend on $\alpha$, only the copula component contributes:
\[
H_{\alpha\mu}
=\nabla_\mu\left(\frac{\partial\cL}{\partial\alpha}\right).
\]
Differentiating term by term, the constant contribution $1/(1+\alpha)$ vanishes. The remaining derivatives are
\[
\nabla_\mu[-\log(V_iV_{i+1})]
=-\frac1{V_i}\nabla_\mu V_i-\frac1{V_{i+1}}\nabla_\mu V_{i+1},
\]
\[
\nabla_\mu\left[\frac1{\alpha^2}\log C_i\right]
=\frac{1}{\alpha^2C_i}\nabla_\mu C_i,
\]
and, by the quotient rule,
\[
\nabla_\mu\left(\frac{\partial_\alpha C_i}{C_i}\right)
=\frac{C_i\nabla_\mu(\partial_\alpha C_i)-(\partial_\alpha C_i)\nabla_\mu C_i}{C_i^2}.
\]
Combining these terms gives
\begin{align}
H_{\alpha\mu}=\sum_{i=1}^{K-1}\Bigg[
&-\frac1{V_i}\nabla_\mu V_i
-\frac1{V_{i+1}}\nabla_\mu V_{i+1}
+\frac{1}{\alpha^2C_i}\nabla_\mu C_i \\
&-\left(2+\frac1\alpha\right)
\frac{C_i\nabla_\mu(\partial_\alpha C_i)-(\partial_\alpha C_i)\nabla_\mu C_i}{C_i^2}
\Bigg].
\end{align}
The mixed derivative in the numerator is obtained from
\[
\partial_\alpha C_i=-V_i^{-\alpha}\log V_i-V_{i+1}^{-\alpha}\log V_{i+1},
\]
which yields
\begin{align}
\nabla_\mu(\partial_\alpha C_i)
={}&V_i^{-\alpha-1}(\alpha\log V_i-1)\nabla_\mu V_i \\
&+V_{i+1}^{-\alpha-1}(\alpha\log V_{i+1}-1)\nabla_\mu V_{i+1}.
\end{align}
The gradients $\nabla_\mu V_i$ and $\nabla_\mu C_i$ are those defined in Section C.3.

% \subsection{Cross-Derivatives: Copula Parameter $\alpha$ and Cholesky Factor $\mathbf{L}$}

\subsection[Cross-Derivatives: Copula Parameter alpha and Cholesky Factor L]{Cross-Derivatives: Copula Parameter $\alpha$ and Cholesky Factor ${L}$}

The mixed derivative involving $\alpha$ and $L$ is obtained by differentiating the scalar score
\[
S_\alpha=\frac{\partial L_c}{\partial\alpha}
\]
with respect to $L$. From Section C.4.1, for any scalar function depending on $z_i$,
\[
\nabla_LS_\alpha
=-L^{-\T}\sum_{i=1}^{K}\delta_i^\alpha z_i^{\T},
\qquad
\delta_i^\alpha=\nabla_{z_i}S_\alpha.
\]
Because $V_i$ is scalar while $z_i$ is vector-valued, the chain rule gives
\[
\delta_i^\alpha
=\frac{\partial S_\alpha}{\partial V_i}\nabla_{z_i}V_i.
\]
Define
\[
\eta_i=\frac{\partial S_\alpha}{\partial V_i}.
\]
Since $L_c$ is a sum over adjacent pairs $k=1,\ldots,K-1$, the variable $V_i$ appears in the $k=i-1$ term as the second argument and in the $k=i$ term as the first argument. Hence
\[
\eta_i=
\sum_{k\in\{i-1,i\}\cap\{1,\ldots,K-1\}}
\left[
-\frac1{V_i}
+\frac{1}{\alpha^2C_k}\frac{\partial C_k}{\partial V_i}
-\left(2+\frac1\alpha\right)
\frac{C_k\,\partial_{\alpha V_i}^2C_k-(\partial_\alpha C_k)(\partial_{V_i}C_k)}{C_k^2}
\right].
\]
The required mixed derivative is
\[
\partial_{\alpha V_i}^2C_i
=V_i^{-\alpha-1}(\alpha\log V_i-1),
\]
with the analogous expression when $V_i$ enters $C_{i-1}$ as its second argument.

Finally, using
\[
\nabla_{z_i}V_i=f_{R_i}(R_i)m(z_i),
\]
the cross-derivative matrix becomes
\[
\boxed{
H_{L\alpha}
=-L^{-\T}\sum_{i=1}^{K}
\left[\eta_i f_{R_i}(R_i)m(z_i)\right]z_i^{\T}.
}
\]

\subsection[Cross-Derivatives: Cholesky Factor L and Mean Vector mu]{Cross-Derivatives: Cholesky Factor $ {L}$ and Mean Vector $ {\mu}$}

The mixed derivative matrix $H_{L\mu}$ characterizes the sensitivity of the gradient with respect to $L$ to perturbations in the mean vector $\mu$. More precisely,
\[
H_{L\mu}
=\nabla_\mu\left(\nabla_L\cL\right)^{\T}
%\frac{\partial}{\partial\mu^{\T}}
%\vecop\left(\frac{\partial\cL}{\partial L}\right)
\in\R^{d\times d^2}.
\]
From Section C.4,
\[
G=L^{-\T}\Omega_m-L^{-\T}\Omega_c.
\]
When differentiating with respect to $\mu$, $L$ is held fixed, so
\[
d_\mu(L^{-\T})=0,
\qquad
d_\mu z_i=-L^{-1}d\mu.
\]
The marginal and copula contributions can therefore be treated separately.
\subsubsection{Marginal Contribution}
For
\[
L^{-\T}\Omega_m
=L^{-\T}\left(\sum_{i=1}^{K}z_i z_i^{\T}-KI_d\right),
\]
we have
\[
d_\mu(L^{-\T}\Omega_m)
=\sum_{i=1}^{K}L^{-\T}\left[(d_\mu z_i)z_i^{\T}+z_i(d_\mu z_i)^{\T}\right].
\]
Substituting $d_\mu z_i=-L^{-1}d\mu$ gives
\[
d_\mu(L^{-\T}\Omega_m)
=\sum_{i=1}^{K}\left[-L^{-\T}L^{-1}(d\mu)z_i^{\T}
-L^{-\T}z_i(d\mu)^{\T}L^{-\T}\right].
\]
After vectorization, the marginal contribution is
\[
\boxed{
T_m=-\sum_{i=1}^{K}\left[
 z_i\otimes L^{-\T}L^{-1}
+L^{-1}\otimes L^{-\T}z_i
\right].
}
\]

\subsubsection{Copula Contribution}
For the copula term,
\[
-L^{-\T}\Omega_c
=-L^{-\T}\left(\sum_{i=1}^{K}\delta_i z_i^{\T}\right),
\]
and therefore
\[
d_\mu(-L^{-\T}\Omega_c)
=-L^{-\T}\sum_{i=1}^{K}
\left[(d_\mu\delta_i)z_i^{\T}+\delta_i(d_\mu z_i)^{\T}\right].
\]
Using
\[
d_\mu z_i=-L^{-1}d\mu,
\qquad
d_\mu\delta_i=\sum_jJ_{ij}(-L^{-1}d\mu),
\]
the first contribution, arising from $\delta_i(d_\mu z_i)^{\T}$, becomes
\[
T_{c1}=\sum_{i=1}^{K}\left(L^{-1}\otimes L^{-\T}\delta_i\right).
\]
The contribution generated by $d_\mu\delta_i$ is separated into diagonal and off-diagonal components:
\[
T_{c2}=\sum_{i=1}^{K}
\left(z_i\otimes L^{-\T}J_{ii}L^{-1}\right),
\]
\[
T_{c,\mathrm{cross}}
=\sum_{i\neq j}
\left(z_i\otimes L^{-\T}J_{ij}L^{-1}\right).
\]

\subsubsection{Total Cross-Hessian}
Combining the marginal and copula contributions yields
\[
\boxed{
H_{L\mu}=T_m+T_{c1}+T_{c2}+T_{c,\mathrm{cross}}.
}
\]

\subsection{Assembly of the Full Hessian Matrix}

The full Hessian matrix summarizes the curvature of the log-likelihood with respect to the complete parameter vector
\[
\theta=
\begin{bmatrix}
\mu\\
\vecop(L)\\
\alpha
\end{bmatrix}
\in\R^{d+d^2+1}.
\]
For vector-valued parameter groups $\zeta_1$ and $\zeta_2$, by defining 
\[
H_{\zeta_1\zeta_2}
=\nabla_{\zeta_1}\left(\nabla_{\zeta_2}\cL\right)^{\T}
\in\R^{\mathrm{dim}(\zeta_1)\times \mathrm{dim}(\zeta_2)},
\]
it is represented as the symmetric block matrix
\[
H=\nabla_{\Theta}\left(\nabla_{\Theta^{\T}}\cL\right)^{\T}
=\begin{bmatrix}
H_{\mu\mu} & H_{\mu L} & H_{\mu\alpha}\\
H_{L\mu} & H_{LL} & H_{L\alpha}\\
H_{\alpha\mu} & H_{\alpha L} & H_{\alpha\alpha}
\end{bmatrix}.
\]
where the dimensions and sources of the individual blocks are summarized as follows
\begin{enumerate}[label=\arabic*.]
\item \textbf{Diagonal blocks}
\begin{align*}
H_{\mu\mu}&\in\R^{d\times d}, &&\text{derived in Section C.3},\\
H_{LL}&\in\R^{d^2\times d^2}, &&\text{derived in Section C.4},\\
H_{\alpha\alpha}&\in\R^{1\times1}, &&\text{derived in Section C.2}.
\end{align*}

\item \textbf{Off-diagonal blocks (cross-derivatives)}
\begin{align*}
H_{L\mu}&\in\R^{d \times d^2}, &&\text{derived in Section C.7},
& H_{\mu L}&=H_{L\mu}^{\T},\\
H_{L\alpha}&\in\R^{1 \times d^2}, &&\text{derived in Section C.6},
& H_{\alpha L}&=H_{L\alpha}^{\T},\\
H_{\alpha\mu}&\in\R^{d\times 1}, &&\text{derived in Section C.5},
& H_{\mu\alpha}&=H_{\alpha\mu}^{\T}.
\end{align*}
\end{enumerate}

This full Hessian matrix is central to both optimization and inference. It provides the second-order information required by Newton--Raphson-type algorithms and, when evaluated at the maximum-likelihood estimator, can be used to construct the observed Fisher information matrix, approximate covariance matrix, and standard errors of the parameter estimates.

\end{appendix}

 \bibliographystyle{plainnat}
 \bibliography{VC090826}

@book{harville1997,
  author    = {Harville, D. A.},
  title     = {Matrix Algebra From a Statistician's Perspective},
  publisher = {Springer},
  year      = {1997}
}

@book{lutkepohl2005new,
  author    = {L{\"{u}}tkepohl, H.},
  title     = {New introduction to multiple time series analysis},
  publisher = {Springer Science {{\&}} Business Media},
  year      = {2005}
}

@book{magnus1999,
  author    = {Magnus, J. R. and Neudecker, H.},
  title     = {Matrix Differential Calculus with Applications in Statistics and Econometrics},
  publisher = {John Wiley {{\&}} Sons},
  year      = {1999}
}

@article{mcneil2009,
  author    = {McNeil, A. J. and Ne{\v{s}}lehov{\'{a}}, J.},
  title     = {Multivariate {Archimedean} copulas, $d$-monotone functions and $l_1$-norm symmetric distributions},
  journal   = {The Annals of Statistics},
  volume    = {37},
  number    = {5B},
  pages     = {3059--3097},
  year      = {2009},
  publisher = {Institute of Mathematical Statistics}
}

@manual{quantmod,
  author    = {Ryan, J. A. and Ulrich, J. M.},
  title     = {{quantmod}: Quantitative Financial Modelling Framework},
  year      = {2024},
  note      = {{R} package version 0.4.26},
  url       = {https://CRAN.R-project.org/package=quantmod}
}

@manual{r_core,
  author       = {R-Core-Team},
  title        = {{R}: A Language and Environment for Statistical Computing},
  organization = {{R} Foundation for Statistical Computing},
  address      = {Vienna, Austria},
  year         = {2024},
  url          = {https://www.R-project.org/}
}

@article{sims1980macroeconomics,
  author    = {Sims, C. A.},
  title     = {Macroeconomics and reality},
  journal   = {Econometrica: Journal of the Econometric Society},
  volume    = {48},
  number    = {1},
  pages     = {1--48},
  year      = {1980}
}

@article{sklar1959,
  author    = {Sklar, A.},
  title     = {Fonctions de r{\'{e}}partition {\`{a}} n dimensions et leurs marges},
  journal   = {Publications de l'Institut de Statistique de l'Universit{\'{e}} de Paris},
  volume    = {8},
  pages     = {229--231},
  year      = {1959}
}

@book{tsay2013multivariate,
  author    = {Tsay, R. S.},
  title     = {Multivariate Time Series Analysis: With {R} and Financial Applications},
  year      = {2013},
  publisher = {John Wiley \& Sons},
  address   = {Hoboken, New Jersey}
}

@article{Brechmann2014,
  author		= "Brechmann, E. C.",
  title			= "Hierarchical {K}endall copulas: Properties and inference",
  journal		= "Canadian Journal of Statistics",
  volume		= "42",
  number		= "1",
  pages			= "78—108",
  year			= "2014"
}

@article{CF2006a,
  author		= "Chen, X. and Fan, Y.",
  title			= "Estimation and model selection of semiparametric copula-based multivariate dynamic models under copula misspecification",
  journal		= "Journal of Econometrics",
  volume		= "135",
  number		= "1–2",
  pages			= "125—154",
  year			= "2006"
}

@article{CF2006b,
  author		= "Chen, X. and Fan, Y.",
  title			= "Estimation of copula-based semiparametric time series models",
  journal		= "Journal of Econometrics",
  volume		= "130",
  number		= "2",
  pages			= "307—335",
  year			= "2006"
}

@article{CWY2009,
  author		= "Chen, X. and Wu, W. B. and Yi, Y.",
  title			= "Efficient estimation of copula-based semiparametric {M}arkov models",
  journal		= "The Annals of Statistics",
  volume		= "37",
  number		= "6B",
  pages			= "4214–4253",
  year			= "2009"
}

@article{CopulaMarkov,
  author		= "Darsow, W.F. and Nguyen, B. and Olsen, E.T.",
  title			= "Copulas and {M}arkov processes",
  journal		= "Illinois Journal of Mathematics",
  volume		= "36",
  number		= "4",
  pages			= "600—642",
  year			= "1992"
}

@article{DGP2009,
  author		= "Domma, F. and Giordano, S. and  Perri, P.E. ",
  title			= "Statistical modeling of temporal dependence in financial data via a copula function",
  journal		= "Communications in Statistics-Simulation and Computation",
  volume		= "38",
  number		= "4",
  pages			= "703—728",
  year			= "2009"
}

@article{DMTW2025,
  author		= "Duker, M.-C. and Matteson, D. S. and Tsay, R. S. and Wilms, I.",
  title			= "Vector AutoRegressive Moving Average Models: A Review",
  journal		= "Wiley Interdisciplinary Reviews: Computational Statistics",
  volume		= "17",
  number		= "1",
  pages			= "e70009",
  year			= "2025"
}

@article{Markov02,
  author		= "Emura, T. and Long, T.H. and Sun, L.H.",
  title			= "Routines for performing estimation and statistical process control under copula-based time series models",
  journal		= "Communications in Statistics - Simulation and Computation",
  volume		= "46",
  number		= "4",
  pages			= "3067—3087",
  year			= "2017"
}

@article{FH2023,
  author		= "Fan, Y. and Henry, M.",
  title			= "Vector copulas",
  journal		= "Journal of Econometrics",
  volume		= "234",
  number		= "1",
  pages			= "128—150",
  year			= "2023"
}

@article{Markov03,
  author		= "Huang, X.W. and Emura, T.",
  title			= "Diagnostic procedures for copula-based {M}arkov chain models
for statistical process control",
  journal		= "Communications in Statistics - Simulation and Computation",
  volume		= "50",
  number		= "8",
  pages			= "2345—2367",
  year			= "2021"
}

@article{Huang2021,
  author		= "Huang, X.W. and Wang, W. and Emura, T.",
  title			= "A copula-based {M}arkov chain model for serially dependent event times with a dependent terminal event",
  journal		= "Japanese Journal of Statistics and Data Science",
  volume		= "4",
  number		= "",
  pages			= "917–951",
  year			= "2021"
}

@book{Joe1997,
  author		= "Joe, H.",
  title			= "Multivariate models and dependence",
  publisher		= "Chapman \& Hall",
  year			= "1997"
}

@article{Kim2021,
  author		= "Kim, J.-M. and Baik, J. and Reller, M.",
  title			= "Control charts of mean and variance using copula {M}arkov {SPC} and conditional distribution by copula",
  journal		= "Communications in Statistics - Simulation and Computation",
  volume		= "50",
  number		= "1",
  pages			= "85—102",
  year			= "2021"
}

@article{Li2019,
  author		= "Li, F. and Tang, Y. and Wang, H. J. ",
  title			= "Copula-based semiparametric analysis for time series data with detection limits",
  journal		= "Canadian Journal of Statistics",
  volume		= "47",
  number		= "3",
  pages			= "438—454",
  year			= "2019"
}

@book{SHAKE2020,
  author		= "Sun, L. H. and Huang, X. W. and Alqawba, M. S. and Kim, J. M. and Emura, T.",
  title			= "Copula-Based {M}arkov Models for Time Series: Parametric Inference and Process Control",
  publisher		= "Springer",
  address = "New York",
  year			= "2020"
  }

@article{Zhang2021,
  author		= "Zhang, S. and Zhang, Q.M. and Lin, H.",
  title			= "Goodness-of-fit test of copula functions for semi-parametric univariate time series models",
  journal		= "Statistical Papers",
  volume		= "62",
  number		= "3",
  pages			= "1697–1721",
  year			= "2021"
}

@article{SD2026,
  author		= "Smith, S. M. and Deng, L.",
  title			= "Vector Vine Copula Models for Multivariate Longitudinal Data",
  journal		= "arXiv:2609.19547",
  year			= "2026"
}

\end{document}